# Monotone data augmentation algorithm for longitudinal continuous, binary and ordinal outcomes: a unifying approach

Yongqiang Tang, yongqiang_tang@yahoo.com

Sep 28, 2026

**Abstract**

The monotone data augmentation (MDA) algorithm has been widely used to impute missing data for longitudinal continuous outcomes. Compared to a full data augmentation approach, the MDA scheme accelerates the mixing of the Markov chain, reduces computational costs per iteration, and aids in missing data imputation under nonignorable dropouts. We extend the MDA algorithm to the multivariate probit (MVP) model for longitudinal binary and ordinal outcomes. The MVP model assumes the categorical outcomes are discretized versions of underlying longitudinal latent Gaussian outcomes modeled by a mixed effects model for repeated measures. A parameter expansion strategy is employed to facilitate posterior sampling of the correlation matrix, and expedite the convergence of the Markov chain in MVP. The method enables the sampling of the regression coefficients and covariance matrix for longitudinal continuous, binary and ordinal outcomes in a unified manner. This property aids in understanding the algorithm and developing computer codes for MVP. We also introduce independent Metropolis-Hasting samplers to handle complex priors, and evaluate how the choice between flat and diffuse normal priors for regression coefficients influences parameter estimation and missing data imputation. Numerical examples are used to illustrate the methodology. Sample SAS code is provided in the Supporting Information.

## 1 Introduction

Recent regulatory documents [1, 2] emphasize the importance of prespecifying estimand strategies that include a primary analysis and appropriate sensitivity analyses under different assumptions about the missing data mechanism. Over the past decade, there has been a notable increase in the use of analyses under the missing not at random (MNAR) assumption in late-phase clinical trials, as requested by regulatory agencies. In such analyses, it is often assumed that, following treatment discontinuation, subjects in the experimental treatment group may lose some or all of the benefits gained while on treatment, and missing data are imputed on the basis of pattern mixture models [3, 4, 5]. The purpose of this paper is to extend monotone data augmentation (MDA) based imputation methods [6, 7], originally developed for longitudinal continuous outcomes, to longitudinal binary and ordinal outcomes, thereby providing a unified framework for handling these types of endpoints.

A two-step heuristic approach is often used for missing data imputation in clinical trials. First, Schafer's [8] MDA procedure, designed for incomplete multivariate normal data and implemented in Statistical Analysis Software (SAS), is employed to impute intermittent missing data that occur before the last visit with an observed outcome, even when dealing with categorical data. Second, missing data following dropout or intercurrent events are imputed sequentially by visit [2].

This heuristic approach is popular because it is easy to understand and can be implemented using existing SAS procedures, although customized programming is still required. It can handle both missing at random (MAR) and MNAR mechanisms. Under MAR, any systematic difference between missing and observed values among subjects on the same treatment is assumed to be entirely explained by historical observed data. The dropout missing data are imputed using the data distribution of subjects on the same treatment. If such difference can not be fully predicted by historical information, the mechanism is MNAR or nonignorable. For example, an MNAR mechanism may arise when treatment is discontinued due to safety reasons or lack of efficacy. In two common analyses requested by regulatory agencies, the missing data for subjects who discontinue the experimental treatment are assumed to either follow the response pattern observed in the control group or exhibit a mean response deterioration compared to those who remain on the experimental treatment [3, 4, 5].

Tang [6, 7] presents a Markov chain Monte Carlo (MCMC) algorithm for the mixed-effects model for repeated measures (MMRM) based on the MDA scheme. A key idea is that, the marginal distribution of the observed data in PMMs is identical to that under MAR. Consequently, model parameters generated from MCMC for MMRM using observed data can be directly used to impute post-dropout missing data under both MAR and MNAR. While similar in concept, Tang's method differs subtly from the two-step procedure. The two-step procedure assumes that the baseline covariates and treatment status are normally distributed, uses different default uninformative priors at each step (potentially facing issues with multicollinearity), and samples the model parameters in both steps, leading to longer runtime. Tang's approach streamlines this process by avoiding unnecessary distributional assumptions and eliminating the need for dual prior specifications and samplings of the model parameters. Additionally, Tang's method addresses multicollinearity by flexible specification of the prior distribution, and can handle more complicated MNAR mechanisms, such as the jump-to-reference (J2R), which is formulated on the basis of the marginal response distribution instead of the sequential regression. Tang [9] also proposes a simple way to implement MNAR imputation using the SAS MI procedure, in which the missing data under MNAR dropout are obtained by simply subtracting the mean difference, calculated from the posterior draw of the model parameters, from the imputed MAR values at each MCMC iteration.

| | Prior | | |
|---|---|---|---|
| Method | covariance matrix$^{(a)}$ | regression coefficient$^{(b)}$ | Comments |
| Liu [10] | IW or Jeffrey | flat | (1) no covariate, or normally distributed covariates<br>(2) need to arrange longitudinal data in reverse-time order |
| Schafer [8] | Jeffrey | flat | (1) no covariate, or normally distributed covariates<br>(2) potential issues with multicollinearity |
| Tang [6, 7] | IW or Jeffrey | conditionally normal or flat | (1) any types of covariates.<br>(2) consistent with MMRM model commonly used in clinical trials |

Table 1: Comparison of three monotone data augmentation (MDA) algorithms for longitudinal continuous outcomes: $^{(a)}$ For the covariance matrix $\Sigma$, Jeffreys' prior is a degenerate case of the Inverse-Wishart (IW) prior; $^{(b)}$ For the regression coefficient (including intercept), flat prior is a degenerate case of the conditionally normal prior given $\Sigma$.

The MDA algorithm integrates out the post-dropout missing data from the posterior distribution using the blocked and collapsed sampling techniques [11, 12]. This results in faster convergence, reduced computational cost per MCMC iteration, and enhanced capability for handling MNAR imputations [12, 8, 9]. In contrast, a full data augmentation (FDA) scheme requires the post-dropout missing data to first be imputed under MAR during posterior sampling of the model parameters and then be re-imputed under MNAR for the final analysis. The efficiency of the MDA algorithm relative to the FDA approach can be formally justified by the theoretical results of Liu [11] and Liu *et. al.* [12]. Table 1 summarizes three MDA algorithms for the longitudinal continuous outcomes. The algorithms by Liu [10] and Schafer [8] are designed for handling multivariate normal data, where the covariates, if present, are assumed to be normally distributed. Liu's approach relies on Bartlett's decomposition, and organizes longitudinal data in a time-reverse order [6]. Both Schafer's method and Tang's [6, 7] method are built on sequential regression. Schafer's method employs noninformative priors, and may face challenges with multicollinearity (refer to Appendix A.3 for details). Tang's MDA algorithm can be viewed as an extension of Schafer's method, and allow for both informative and uninformative / flat priors within the framework of conjugate matrix-normal-inverse-Wishart (MNIW) priors on the covariance matrix and regression coefficients. It allows any type of covariates and its assumption is consistent with that in MMRM commonly used in the analysis of clinical trials.

Tang [13, 14] extends the MDA algorithm to longitudinal categorical outcomes or mixed-type outcomes. It employs the generalized linear model to depict how the response at each visit relates to the baseline covariates and the response from previous visits. Tang [13] also evaluates a two-step approach, whereby intermittent missing data are imputed via fully conditional specification [15], and post-dropout outcomes are then imputed using sequential logistic regression. Another popular method for longitudinal binary and ordinal endpoints is the multivariate probit (MVP) model [16]. In MVP, the longitudinal categorical outcomes are assumed to be discretized versions of underlying longitudinal latent Gaussian outcomes. Several MCMC algorithms utilizing the FDA scheme have been suggested for MVP. Please refer to Lu and Chen [17] for a review and relevant references.

Bayesian inference is driven by the posterior distribution, formed as the product of the likelihood from the MVP model and the prior density. These MCMC algorithms are expected to yield similar inference results in the analysis of clinical trials when using non-informative or weakly informative priors. However, in the implementation by Lu and Chen, these methods produce somewhat different results. Some issues in Lu and Chen's implementation will be discussed in Sections 4.3 and 6.

We present new MDA algorithms for MVP that allows more flexible and complex prior specifications. We adopt a parameter expansion (PX) strategy [18, 19] to simplify the sampling of correlation matrix and expedite the convergence of the Markov chain. A Gibbs approach is introduced on the basis of a specific yet broadly applicable prior. The posterior distribution for the regression coefficients and covariance matrix in MVP is designed to mirror that of Tang [6, 7] for longitudinal continuous outcomes. It facilitates the understanding of the new MDA scheme and aids in the development of computer codes. The new MDA scheme represents an improvement over our earlier MDA algorithm [13] by retaining its computational advantages while yielding a posterior distribution that is easier to interpret. The new scheme is therefore highly recommended. Additionally, we introduce independent Metropolis-Hasting (iMH) samplers to handle complex priors. While previous work has focused primarily on binary outcomes [17], the proposed MDA scheme is also well suited for longitudinal ordinal outcomes.

The paper is organized as below. In Section 2, we review the MDA scheme for longitudinal continuous outcomes. Section 3 introduces the Gibbs-MDA algorithm for longitudinal binary and ordinal outcomes. Numerical examples are presented in Section 4. We discuss the use of iMH samplers to handle complex priors in Section 5. The appendix provides technical details and compares the posterior distributions obtained under flat and diffuse normal priors on the regression coefficients in MMRM.

# 2 MDA scheme for MMRM: review

We review the MDA algorithm for MMRM. The key idea is to factorize the MMRM into a sequence of conditional regressions, and decompose the conjugate MNIW prior into independent conjugate normal–gamma (NG) priors for these sequential regressions. These priors can be specified to be either degenerate or weakly informative.

| Monotone data structure | | | | | | | | |
|---|---|---|---|---|---|---|---|---|
| subject | $x_1$ | ... | $x_q$ | $y_1$ | ... | $y_{p-1}$ | $y_p$ | |
| 1 | $x_{11}$ | ... | $x_{1q}$ | $y_{11}$ | ... | $y_{1,p-1}$ | $y_{1p}$ | pattern $p$ |
| ⋮ | ⋮ | ... | ... | ... | | ⋮ | ⋮ | |
| $n_p$ | $x_{n_p1}$ | ... | $x_{n_pq}$ | $y_{n_p1}$ | ... | $y_{n_p,p-1}$ | $y_{n_pp}$ | |
| $n_p+1$ | $x_{n_p+1,1}$ | ... | $x_{n_p+1,q}$ | $y_{n_p+1,1}$ | ... | $y_{n_p+1,p-1}$ | | potential pattern $p-1$ |
| ⋮ | ⋮ | ... | ... | ... | ... | ⋮ | | |
| $n_{p-1}$ | $x_{n_{p-1},1}$ | ... | $x_{n_{p-1},q}$ | $y_{n_{p-1},1}$ | ... | $y_{n_{p-1},p-1}$ | | |
| ⋮ | ⋮ | ... | ... | ⋮ | | | | |
| $n_2+1$ | $x_{n_2+1,1}$ | ... | $x_{n_2+1,q}$ | $y_{n_2+1,1}$ | | | | potential pattern 1 |
| ⋮ | ... | ... | | | | | | |
| $n_1$ | $x_{n_11}$ | ... | $x_{n_1q}$ | $y_{n_11}$ | | | | |
| ⋮ | ... | ... | | | | | | potential pattern 0 |
| $n$ | $x_{n1}$ | ... | $x_{nq}$ | | | | | |

Table 2: Monotone data structure: Subject $i$ is in pattern $s$ if $y_{is}$ is the last observed outcome. Subjects are ordered by pattern in descending order. At visit $j$, $n_j$ subjects from patterns $(j, \dots, p)$ have non-missing $y_{ij}$, with $n_p \leq \dots \leq n_1 \leq n$. For complete data, all subjects are in pattern $p$, so $n_p = \dots = n_1 = n$. Multivariate normal data are a special case with one covariate ($q = 1$, $x_{i1} \equiv 1$).

## 2.1 Data structure

Suppose we collect data at $p$ fixed post-baseline visits in a clinical trial. Let $\mathbf{y}_i = (y_{i1}, \dots, y_{ip})'$ be the outcomes at the $p$ visits, and $\mathbf{x}_i = (x_{i1}, \dots, x_{iq})'$ the baseline covariates (including intercept and treatment status) for subject $i$. Let $s_i$ index the dropout pattern for subject $i$. Subject $i$ is in pattern $s$ if $y_{is}$ is the last observed measurement, and in pattern 0 if the subject has no post-baseline assessment. Without loss of generality, we sort the data so that subjects in pattern $s$ are arranged before subjects in pattern $t$ if $s > t$. Table 2 shows the monotone data pattern, in which there are no missing data before the last observed value for each subject. Suppose $y_{ik}$ is not missing in the first $n_k$ subjects. For complete data, all subjects are in pattern $p$ with $n_1 = \dots = n_p = n$.

Let $\mathbf{y}_{io}$, $\mathbf{y}_{ii}$ and $\mathbf{y}_{id}$ be the observed, intermittent missing, and post-dropout outcomes, respectively, for subject $i$. Let $\mathbf{y}_{im} = (y_{i1}, \dots, y_{is_i})' = \{\mathbf{y}_{io}, \mathbf{y}_{ii}\}$ be the outcomes prior to dropout. Let $Y_{\text{o}}$, $Y_{\text{I}}$ and $Y_{\text{d}}$ be the observed, intermittent missing, and post-dropout outcomes from all subjects, respectively. Then $Y_{\text{m}} = (Y_{\text{O}}, Y_{\text{I}})$ represents the monotone data in Table 2, and $Y = (Y_{\text{o}}, Y_{\text{I}}, Y_{\text{d}})$ is the full data.

For simplicity, the term *dropout* is used broadly to include intercurrent events [2]. The handling of missing data depends on the estimand strategy [2]. For illustration, consider a clinical trial in which some subjects may use rescue medication, potentially confounding treatment comparisons. Suppose we are interested in a hypothetical estimand reflecting the treatment effect if subjects had continued the randomized treatment without using rescue medication. In this setting, outcomes following treatment discontinuation or the use of rescue medication are considered missing. In sensitivity analyses or even in the primary analysis, the missing data may be assumed to follow a common MAR or MNAR mechanism, or mechanisms that vary by the intercurrent event type or dropout reason.

## 2.2 MMRM is factorized as a sequence of conditional regression

The MMRM is widely used to analyze longitudinal continuous outcomes in clinical trials [20]

$$\begin{cases} y_{i1} & = \alpha_{11}x_{i1} + ... + \alpha_{1q}x_{iq} + \varepsilon_{i1} \\ & \quad ... \\ y_{ip} & = \alpha_{p1}x_{i1} + ... + \alpha_{pq}x_{iq} + \varepsilon_{ip} \end{cases} \tag{1}$$

where $(\varepsilon_{i1}, ..., \varepsilon_{ip})' \sim N(\mathbf{0}, \Sigma)$ for an unstructured covariance matrix $\Sigma$. The preference for the MMRM in the analysis of clinical trials arises from its minimal assumptions about the response trajectory over time and the covariance structure for within-subject errors [20].

The MMRM can be factorized as a sequence of linear regressions of $y_{ij}$ on the baseline covariates $(x_{i1}, ..., x_{iq})$ and historical outcomes $(y_{i1}, ..., y_{i,j-1})$

$$y_{ij} = \tilde{\alpha}_{j1}x_{i1} + ... + \tilde{\alpha}_{jq}x_{iq} + \beta_{j1}y_{i1} + ... + \beta_{j,j-1}y_{i,j-1} + \epsilon_{ij} \tag{2}$$

for $j = 1, ..., p$, where $(\epsilon_{i1}, ..., \epsilon_{ip})' \sim N(\mathbf{0}, \Lambda)$ and $\Lambda = \text{diag}(\gamma_1^{-1}, ..., \gamma_p^{-1})$.

The covariance matrix $\Sigma$ in MMRM and the parameters $(\beta_{ij}, \gamma_j)$ in the sequential regression are related through the LDL decomposition $\Sigma = L\Lambda L'$, where $U = \begin{bmatrix} 1 & 0 & ... & 0 \\ -\beta_{21} & 1 & ... & 0 \\ & ... & ... & 0 \\ -\beta_{p1} & ... & -\beta_{p,p-1} & 1 \end{bmatrix}$ and $L = U^{-1}$.

Furthermore, $\tilde{\alpha} = \begin{bmatrix} \tilde{\alpha}_{11} & ... & \tilde{\alpha}_{1q} \\ \vdots & ... & \vdots \\ \tilde{\alpha}_{p1} & ... & \tilde{\alpha}_{pq} \end{bmatrix} = U\alpha$ and $\alpha = L\tilde{\alpha}$, where $\alpha = \begin{bmatrix} \alpha_{11} & ... & \alpha_{1q} \\ \vdots & ... & \vdots \\ \alpha_{p1} & ... & \alpha_{pq} \end{bmatrix}$.

## 2.3 Conjugate MNIW prior is decomposed as independent NG priors

Tang [6] considers a conjugate MNIW prior for $(\Sigma, \alpha)$, denoted by $MNIW(A, \nu_0, \alpha_0, M)$. Specifically, an inverse-Wishart prior $IW(A, \nu_0)$ is put on $\Sigma$. Setting $(\nu_0 = 0, A = \mathbf{0})$ yields Jeffrey's prior for $\Sigma$. The conditional distribution of $\alpha$ given $\Sigma$ is the possibly degenerate matrix normal $\alpha|\Sigma \sim MN(\alpha_0, \Sigma, M^+)$, where the column covariance matrix $M^+$ is Moore-Penrose inverse of the $q \times q$ precision matrix $M$ with rank $r$, where $r$ is the number of covariates with non-flat priors

$$\begin{gathered} \Sigma \sim IW(A, \nu_0) \\ \alpha|\Sigma \sim MN(\alpha_0, \Sigma, M^+) \end{gathered} \tag{3}$$

A unique feature of the prior specification is its flexibility in assigning flat priors to any of the $q$ covariates. When $M = \mathbf{0}$ (i.e. $M^+ = \mathbf{0}$), all covariates have flat priors. We may assign weakly to moderately informative prior to certain covariates while setting flat prior on others. This scenario may arise when historical information is available for the control treatment but not for the experimental treatment. If all entries in the $k$th row and $k$th column of $M$ are 0, the prior on the regression coefficients $(\alpha_{1k}, \dots, \alpha_{pk})'$ for $x_{ik}$ is flat. Consequently, the prior mean for $(\alpha_{1k}, \dots, \alpha_{pk})'$ is automatically set to $\mathbf{0}$ as it has no impact on the prior density. The conditional prior density of $\alpha$ given $\Sigma$ is given by

$$p(\alpha|\Sigma) \propto |\Sigma|^{-r/2} \exp\left[-\text{tr}(M(\alpha-\alpha_0)'\Sigma^{-1}(\alpha-\alpha_0))/2\right]$$

As an illustration, we consider the multivariate normal distribution. It is a special case of the MMRM with one covariate ($q = 1$, $x_{i1} \equiv 1$). The prior for $\alpha = \mu = (\mu_1, \dots, \mu_p)'$ given $\Sigma$ is a possibly degenerate normal distribution $N(\mu_0, m_0^+\Sigma)$ with density function

$$p(\mu|\Sigma) \propto |\Sigma|^{-r/2} \exp[-m_0(\mu-\mu_0)\Sigma^{-1}(\mu-\mu_0)/2], \tag{4}$$

It is a unified notation for a normal or flat prior. When $M = m_0 > 0$, it is a normal prior since $m_0^+ = m_0^{-1}$ and $r = I(m_0 > 0) = 1$. When $m_0 = 0$, we get $m_0^+ = 0$, $r = I(m_0 > 0) = 0$, and $\mu$ has a flat prior since $p(\mu|\Sigma) \propto 1$.

In the MDA scheme, we work with the sequential regression (2) instead of directly dealing with model (1). This choice is motivated by the property that the MNIW prior can be decomposed into independent conjugate NG distributions for $(\theta_j, \gamma_j)$ under monotone data pattern [6, 9], where $\theta_j = (\tilde{\alpha}_{j1}, \dots, \tilde{\alpha}_{jq}, \beta_{j1}, \dots, \beta_{j,j-1})'$ is the regression coefficient, and $\gamma_j$ is the precision parameter in the regression model (2) for pattern $j$.

The NG distribution is the conjugate prior for the regression coefficient and precision parameter in univariate linear regression. Its joint probability density function (PDF), denoted by $NG(f, D)$, is

$$f(\theta,\gamma) = \frac{|\Omega|^{1/2}\ (a/2)^{f/2}}{\sqrt{(2\pi)^{(m-1)}}\ \Gamma(a/2)}\, \gamma^{(f+m-1)/2-1} \exp\left(-\gamma\tilde{\theta}' D\tilde{\theta}/2\right), \tag{5}$$

where $m \geq 2$, $\tilde{\theta} = \begin{bmatrix} -\theta \\ 1 \end{bmatrix}$ and $D$ is a $m \times m$ matrix. Writing $D$ as $\begin{bmatrix} \Omega & \Omega\mu \\ \mu'\Omega & a + \mu'\Omega\mu \end{bmatrix}$, the NG distribution factorizes into a marginal gamma (equivalently, chi-square) distribution for the precision parameter $\gamma$ and a conditional normal distribution for the $(m-1) \times 1$ regression coefficient vector $\theta$

$$\begin{aligned} \gamma &\sim GA(f/2, a/2) \text{ (or equivalently } \gamma \sim \chi_f^2/a) \\ \theta|\gamma &\sim N(\mu, \gamma^{-1}\Omega^{-1}) \end{aligned} \tag{6}$$

Although this distribution is well known, using its joint form as specified in Equation (5), rather than the factorized form in Equation (6), greatly facilitate the derivation of the posterior distribution in the MDA algorithm [6], and the development of algorithms under complex priors. Appendix A.1 describes several methods for generating samples from the NG distribution.

Lemma 1 below summarizes the key results of Tang [6, 9] on the prior and posterior NG distribution of $(\theta_j, \gamma_j)$ under the monotone data pattern.

**Lemma 1** *a) The $MNIW(A, \nu_0, \alpha_0, M)$ prior can be factorized as independent NG priors on $(\theta_j, \gamma_j)$'s*

$$\theta_j, \gamma_j \propto NG[f_{j0}, D_{j0}] \tag{7}$$

*where $f_{j0} = \nu_0 + j - p - (q - r)$, and $D_{j0}$ is the leading $(q + j) \times (q + j)$ submatrix of $D_0 = \begin{bmatrix} M & M\alpha_0' \\ \alpha_0 M & \alpha_0 M\alpha_0' + A \end{bmatrix}$.*

*b) Given the monotone data in Table 2, $(\theta_j, \gamma_j)$'s follow independent NG posterior distributions*

$$\theta_j, \gamma_j | Y_m \propto NG[f_j, D_j] \tag{8}$$

*where* $f_j = n_j + f_{j0}$, $D_j = D_{j0} + Z_j' Z_j$ *and* $Z_j = \begin{bmatrix} x_{11} & \cdots & x_{1q} & y_{11} & \cdots & y_{ij} \\ \vdots & & \cdots & & \vdots & \\ x_{n_j 1} & \cdots & x_{n_j q} & y_{n_j 1} & \cdots & y_{n_j j} \end{bmatrix}$.

For multivariate normal data, we set $D_0 = \begin{bmatrix} m_0 & m_0 \mu_0' \\ m_0 \mu_0 & m_0 \mu_0 \mu_0' + A \end{bmatrix}$, $Z_j = \begin{bmatrix} 1 & y_{11} & \cdots & y_{ij} \\ & \cdots & & \\ 1 & y_{n_j 1} & \cdots & y_{n_j j} \end{bmatrix}$ and $q = 1$ under the $IW(A, \nu_0)$ prior on $\Sigma$, and the prior in Equation (4) on $\mu$. Schaffer's MDA algorithm is a special case with Jeffrey's prior for $\Sigma$ and flat prior for $\mu$ (i.e. $m_0 = 0$, $\nu_0 = 0$, $A = \mathbf{0}$).

The MDA algorithm is presented in Algorithm 1. Compared with the FDA approach outlined in Appendix A.2, the MDA typically converges faster and incurs lower per-iteration computational cost.

**Algorithm 1: MDA for longitudinal continuous outcomes**

The MDA algorithm iterates between the following two steps

1. For $i = 1, \ldots, n$, impute $\mathbf{y}_{i\mathrm{i}}$ from the conditional normal distribution $f(\mathbf{y}_{i\mathrm{i}} \mid \mathbf{y}_{io}, \{\theta_j, \gamma_j\}_{j=1}^p)$, as given in Tang [7]. This step creates the monotone structure in Table 2.

2. For $j = 1, \ldots, p$, sample $(\theta_j, \gamma_j)$ from the posterior distribution in Equation (8)

In practice, noninformative priors are often employed when prior information is limited, or when the inference is intended to rely primarily on the observed data. However, Appendix A.3 shows that the posterior distributions under flat and diffuse normal priors on $\alpha$ differ to some extent in MMRM.

# 3 Binary and ordinal outcomes

## 3.1 Multivariate probit models with parameter restriction

The ordinal outcome is modeled using an MVP model, with the binary outcome as a special case. Let $w_{ij} = 1, \ldots, K$ be the ordinal outcome with $K$ levels. For binary endpoints, we have $K = 2$ and code the value as 1 and 2 instead of the traditional 0 and 1.

We introduce the latent variables $\mathring{\mathbf{y}}_i = (\mathring{y}_{i1}, \ldots, \mathring{y}_{ip})' \sim N(\mathring{\alpha}\mathbf{x}_i', \Sigma)$ and the cut points $-\infty = \mathring{c}_{j0} < \mathring{c}_{j1} < \ldots < \mathring{c}_{jK} = \infty$ at visit $j$ for $j = 1, \ldots, p$. The cut points define the thresholds between consecutive categories on the underlying latent variable $y_{ij}$'s. The ordinal outcomes and latent variables are related such that $w_{ij} = k$ if and only if $\mathring{c}_{j,k-1} < \mathring{y}_{ij} \leq \mathring{c}_{jk}$ for $j = 1, \ldots, p$. We set $\mathring{c}_{j1} = 0$ since it will be absorbed into the intercept term. We allow the cut points to vary across visits for ordinal outcomes.

The model parameters are not identifiable. For example, if $(\mathring{y}_{ij}, \mathring{\alpha}_{jk}, \mathring{c}_{jk})$'s are inflated 10 times, and $\Sigma$ is inflated 100 times, we will get the same likelihood for the observed data. We restrict $\Sigma$ to be a correlation matrix $\mathsf{R}$ to ensure parameter identifiability. Other restrictions are possible. For example, Tang [13] constrains the conditional variance of $\mathring{y}_{ij}$ given $(\mathbf{x}_i, \mathring{y}_{i1}, \ldots, \mathring{y}_{ij-1})$ to be 1, and Liu and Daniel [21] assumes $\sum_{i=1}^n (\mathring{y}_{ij} - x_{ij}\mathring{\alpha})^2 = 1$ for $j = 1, \ldots, p$.

Let $\mathring{C}_{ij}$ denote the interval containing the possible value for the latent variable $\mathring{y}_{ij}$. That is, $\mathring{C}_{ij} = (\mathring{c}_{j,k-1}, \mathring{c}_{jk})$ if subject $i$ is in category $k$ at visit $j$, and $\mathring{C}_{ij} = (-\infty, \infty)$ if $w_{ij}$ is missing.

Then $\mathring{\mathbf{y}}_i$ follows the truncated multivarite normal distribution

$$\mathring{\mathbf{y}}_i \sim N(\mathring{\alpha}\mathbf{x}_i', \mathsf{R}) \prod_{j=1}^{p} I(\mathring{y}_{ij} \in \mathring{C}_{ij}).$$

### 3.2 Parameter expanded model

We employ the PX technique by introducing the expansion parameters $\mathbb{D} = \text{diag}(\mathbb{d}_1, \dots, \mathbb{d}_p)$. The latent variables and parameters in the expanded and restricted models are related as

$$y_{ij} = \mathbb{d}_j^{1/2} \mathring{y}_{ij}, \alpha_{jk} = \mathbb{d}_j^{1/2} \mathring{\alpha}_{jk}, c_{jl} = \mathbb{d}_j^{1/2} \mathring{c}_{jl} \text{ and } C_{ij} = \mathbb{d}_j^{1/2} \mathring{C}_{ij}. \quad (9)$$

We have $w_{ij} = k$ if and only if $c_{j,k-1} < y_{ij} \leq c_{jk}$. The transformation preserves the marginal distribution of the observed categorical outcomes $w_{ij}$'s.

Under the expanded model, we have

$$\mathbf{y}_i \sim N(\alpha\mathbf{x}_i', \Sigma) \prod_{j=1}^{p} I(y_{ij} \in C_{ij})$$

For subjects in pattern $s_i = s$, the marginal distributions of the latent outcomes $\mathring{\mathbf{y}}_{im} = (\mathring{y}_{i1}, \dots, \mathring{y}_{is})$ and $\mathbf{y}_{im} = (y_{i1}, \dots, y_{is})$ prior to dropout are truncated multivariate normal

$$\mathring{\mathbf{y}}_{im} \sim N(\mathring{\alpha}_i\mathbf{x}_i', \mathsf{R}_i) \prod_{j=1}^{s} I(\mathring{y}_{ij} \in \mathring{C}_{ij}) \text{ and } \mathbf{y}_{im} \sim N(\alpha_i\mathbf{x}_i', \Sigma_i) \prod_{j=1}^{s} I(y_{ij} \in C_{ij}), \quad (10)$$

where $\mathsf{R}_i$ and $\Sigma_i$ are, respectively, the $s \times s$ leading submatrix of $\mathsf{R}$ and $\Sigma$, and $\mathring{\alpha}_i$ and $\alpha_i$ are, respectively, the matrices containing the first $s$ rows of $\mathring{\alpha}$ and $\alpha$.

The PX model is parameter redundant given the observed data $w_{ij}$'s, but identifiable given the augmented latent variable $y_{ij}$'s. The MVP model is the special case of the PX model at $\mathbb{d}_j = 1$ for $j = 1, \dots, p$, and can be recovered from the PX model by inversely applying the transformation in Equation (9), where $\mathbb{d}_j$ is the $j$th diagonal element of the variance matrix $\Sigma$.

As detailed in Section 3.5, the PX strategy [18, 19] is introduced for two reasons: to accelerate the convergence of MCMC algorithms and to facilitate sampling of the correlation matrix.

### 3.3 The prior

We first consider the following prior for $(\mathsf{R}, \mathring{\alpha}, \mathbf{c})$ in the MVP model

$$\begin{aligned} p(\mathsf{R}) &\propto |\mathsf{R}|^{-(\nu_0+p+1)/2} \prod_{j=1}^{p} \left[\mathsf{R}^{jj}\right]^{-\nu_0/2}, \\ \mathring{\alpha}|\mathsf{R} &\sim MN(\mathbf{0}, \mathsf{R}, M^{-1}) \\ p(\mathbf{c}_j) &\propto N(\mathbf{c}_j|\mathbf{c}_{j0}, V_{cj}) I(0 < c_{j2} < \dots < c_{j,K-1}) \text{ if } K \geq 3 \end{aligned} \quad (11)$$

where $\mathbf{c}_j = (c_{j2}, \dots, c_{j,K-1})'$ and $\mathsf{R}^{jj}$ is the $j$-th diagonal element of $\mathsf{R}^{-1}$. We assume $M$ has full rank, implying that no covariate has a flat prior. The cut points $\mathbf{c}_j$, required only for ordinal outcomes, have a multivariate normal prior with an order constraint $0 < c_{j2} < \dots < c_{j,K-1}$ for $j = 1, \dots, p$. The induced prior for $\mathring{\mathbf{c}}_j$ is a normal mixture given by $\mathring{\mathbf{c}}_j|\mathbb{d}_j, \mathsf{R} \sim N(\mathbb{d}_j^{-1/2}\mathbf{c}_{j0}, \mathbb{d}_j^{-1}V_{cj})$ and $p(\mathring{\mathbf{c}}_j|\mathsf{R}) \propto$

$\int N(\mathring{\mathbf{c}}_j|\mathbb{d}_j^{-1/2}\mathbf{c}_{j0}, \mathbb{d}_j^{-1}V_{cj})p(\mathbb{d}_j|\mathsf{R})d(\mathbb{d}_j)$, where the prior for $\mathbb{d}_j$ is specified later.

When the covariance matrix $\Sigma = \mathbb{D}^{1/2}\mathsf{R}\mathbb{D}^{1/2}$ follows the $IW(\nu_0, I)$ distribution, the prior for the correlation matrix $\mathsf{R}$ is obatined as [22]

$$p(\mathsf{R}) \propto |\mathsf{R}|^{-(\nu_0+p+1)/2} \prod_{j=1}^{p} [\mathsf{R}^{jj}]^{-\nu_0/2}. \tag{12}$$

For any pair $j \neq k$, the marginal distribution of the correlation $r_{jk}$ is a Beta distribution Beta$(\frac{\nu_0-p+1}{2}, \frac{\nu_0-p+1}{2})$ on $[-1, 1]$ with density [22]

$$p(r_{jk}) \propto (1 - r_{jk}^2)^{(\nu_0-p-1)/2}$$

When $\nu_0 = p + 1$, the marginal distribution for any individual correlation is uniform over $[-1, 1]$. When $\nu_0 > p + 1$, the marginal prior density is increasingly concentrated around zero. Conversely, when $\nu_0 < p + 1$, the density places relatively greater weight near the boundaries $\pm 1$.

We specify the prior for the expansion parameters $\mathbb{d}_j$ as

$$\mathbb{d}_j|\mathsf{R} \sim IG(\nu_0/2, a_j\mathsf{R}^{jj}/2) = \frac{a_j\mathsf{R}^{jj}}{\chi^2_{\nu_0}} \tag{13}$$

This induces an $IW(\nu_0, A)$ prior on $\Sigma$ with $A = \text{diag}(a_1, \dots, a_p)$, enabling efficient sampling of $\Sigma$ and $\mathsf{R}$. We use $a_j = 1$. The choice of prior for the expansion parameters is not expected to materially affect inference on $(\mathsf{R}, \mathring{\alpha}, \mathring{\mathbf{c}}_j)$ or the imputation of missing response statuses $w_{ij}$ under a diffuse prior on $\mathbf{c}_j$.

### 3.4 Posterior distribution

Appendix A.4 derives the posterior distribution in MVP. The posterior distribution of $(\Sigma, \alpha)$ has the same expression as that in MMRM under the $MNIW(\nu_0, A, \mathbf{0}, M)$ prior

$$\begin{aligned} &\pi(\Sigma, \alpha|Y_{\text{m}}, W_o) = \psi(\Sigma, \alpha|\nu_0, A, \mathbf{0}, M) \\ &\propto \left\{|\Sigma|^{-(\nu_0+p+1)/2} \exp[-\text{tr}(A\Sigma^{-1})/2]\right\} \left\{|\Sigma|^{-q/2} \exp[-\text{tr}\{M\alpha'\Sigma^{-1}\alpha/2\}]\right\} \prod_{i=1}^{n} N(Y_{i\text{m}}|\alpha_i\mathbf{x}_i', \Sigma_i). \end{aligned} \tag{14}$$

where $A = \text{diag}(a_1, \dots, a_p)$, and $W_o = \{w_{ij} : i = 1, \dots, n, j = 1, \dots, p\}$ denotes the observed response statuses from all subjects. Once the latent variables $Y_{\text{m}}$ are imputed, the MVP model reduces to an MMRM framework. We can employ the same MDA scheme as MMRM to draw $(\theta_j, \gamma_j)$'s, thereby facilitating both understanding of the algorithm and its implementation.

We need to sample $\mathbf{c}_j$ only for ordinal outcomes. Its posterior distribution is truncated normal

$$\pi(\mathbf{c}_j|\Sigma, \alpha, Y_{\text{m}}, W_o) \propto N(\mathbf{c}_j|\,\mathbf{c}_{j0}, V_{cj}) \prod_{k=2}^{K-1} I[\underline{\mathsf{m}}_{jk} \leq c_{jk} < \bar{\mathsf{m}}_{jk}]. \tag{15}$$

where $\underline{\mathsf{m}}_{jk} = \max_{\{i:w_{ij}=k\}}\{y_{ij}\}$ and $\bar{\mathsf{m}}_{jk} = \min_{\{i:w_{ij}=k+1\}}\{y_{ij}\}$. The order constraint $0 < c_{j2} < \dots < c_{j,K-1}$ is automatically ensured in the posterior distribution.

Under a relatively diffuse normal prior on $\mathbf{c}_j$, the truncated posterior in Equation (15) is well approximated by the uniform distribution over its support interval

$$\mathbf{c}_j|\Sigma, \alpha, Y_{\text{m}}, W_o \propto \prod_{k=2}^{K-1} I[\underline{\mathsf{m}}_{jk} \leq c_{jk} < \bar{\mathsf{m}}_{jk}]. \tag{16}$$

We can use this uniform distribution directly as an approximation or as a proposal density to sample from the posterior in Equation (15) via rejection sampling [23]. While this simple approach is typically sufficient, alternative methods can be used when needed. For $K \geq 4$, we can apply the Gibbs sampler of Li and Ghosh [24], whereas for $K = 3$, $\mathbf{c}_j$ is univariate and can be drawn by Robert's method [23].

In practice, a normal prior is preferred to a flat prior, as it ensures that the posterior distribution is proper. However, the posterior distributions are similar under the two priors for the cut points $\mathbf{c}_j$.

### 3.5 Gibbs algorithm

This section first presents the concept of the PX-based MCMC scheme, and then introduces the Gibbs algorithm based on the PX [18, 19] and MDA techniques.

**Conceptual Algorithm 2: MDA for longitudinal binary or ordinal outcomes**

1. Draw from $\pi(\mathsf{R}, \mathring{\alpha}, \mathring{\mathbf{c}} | Y_{\text{m}}, W_o)$: Generate $(\Sigma^{\text{old}}, \alpha^{\text{old}}, \mathbf{c}^{\text{old}})$ from the posterior $\pi(\Sigma, \alpha, \mathbf{c} | Y_{\text{m}}, W_o) \propto \pi(\Sigma, \alpha | Y_{\text{m}}, W_o)\, \pi(\mathbf{c} | \Sigma, \alpha, Y_{\text{m}}, W_o)$, and convert it to $(\mathbb{D}^{\text{old}}, \mathsf{R}, \mathring{\alpha}, \mathring{\mathbf{c}})$

2. Impute $y_{i\text{m}}$ for all subjects: Draw $\mathbb{D}^{\text{new}} = \text{diag}(\mathbb{d}_1^{\text{new}}, \dots, \mathbb{d}_p^{\text{new}})$ from $\mathbb{d}_j | \mathsf{R} \sim IG(\nu_0/2, a_j \mathsf{R}^{jj}/2)$, and impute $\mathbf{y}_{i\text{m}}$ from the truncated normal distribution $f(\mathbf{y}_{i\text{m}} | \mathbb{D} = \mathbb{D}^{\text{new}}, \mathsf{R}, \mathring{\alpha}, \mathring{\mathbf{c}}, W_o)$. The default value of $a_j$ is 1.

*Note:* For ease of exposition, we use $(\Sigma, \alpha)$ as shorthand for $\{(\theta_j, \gamma_j) : j = 1, \dots, p\}$ in the MDA algorithm.

In the restricted model, the parameter space is $(\mathsf{R}, \mathring{\alpha}, \mathring{\mathbf{c}}, \mathring{Y}_{\text{m}})$. A traditional MCMC scheme alternates between drawing from $\pi(\mathsf{R}, \mathring{\boldsymbol{\alpha}}, \mathring{\mathbf{c}} \mid \mathring{Y}_{\text{m}}, W_o)$ and $\pi(\mathring{Y}_{\text{m}} \mid \mathsf{R}, \mathring{\boldsymbol{\alpha}}, \mathring{\mathbf{c}}, W_o)$. In contrast, Algorithm 2 employs a PX scheme with the enlarged parameter space $(\mathsf{R}, \mathring{\alpha}, \mathring{\mathbf{c}}, \mathring{Y}_{\text{m}}, \mathbb{D})$, and iteratively samples from $\pi(\mathsf{R}, \mathring{\alpha}, \mathring{\mathbf{c}} | Y_{\text{m}}, W_o)$ and $\pi(Y_{\text{m}} | \mathsf{R}, \mathring{\alpha}, \mathring{\mathbf{c}}, W_o)$.

The expansion parameters $\mathbb{D}$ are implicitly generated twice per MCMC iteration. In a PX scheme, the expansion parameters are introduced to speed up the convergence of the MCMC algorithm by allowing the Markov chain to move more freely in the expanded parameter space, thereby reducing autocorrelation among posterior samples [18, 19].

The PX scheme also facilitates sampling of the correlation matrix $\mathsf{R}$. Direct sampling of a positive-definite correlation matrix can be difficult. Instead, the PX scheme samples the unstructured covariance matrix $\Sigma = \mathbb{D}^{1/2} \mathsf{R} \mathbb{D}^{1/2}$ and then recover the correlation matrix $\mathsf{R}$ from $\Sigma$. Because $\Sigma$ is positive definite by construction, the resulting correlation matrix $\mathsf{R}$ is guaranteed to be positive definite [25].

Algorithm 2 serves as a conceptual framework, which we adapt for practical implementation using a Gibbs scheme. In Step 1, sampling $(\alpha^{\text{old}}, \Sigma^{\text{old}})$ from $\pi(\Sigma, \alpha | Y_{\text{m}}, W_o)$ is achieved by drawing $(\beta_j, \gamma_j)$'s in the same manner as for continuous outcomes. In Step 2, we update $\mathbb{D}^{\text{old}}$ to $\mathbb{D}^{\text{new}}$, while holding $(\mathsf{R}, \mathring{\alpha}, \mathring{\mathbf{c}})$ fixed. We impute $\mathbf{y}_{i\text{m}}^{\text{old}} \sim f(\mathbf{y}_{i\text{m}} | \Sigma^{\text{old}}, \alpha^{\text{old}}, \mathbf{c}^{\text{old}}, W_o)$, as given in Equation (10), and update $y_{ij} = y_{ij}^{\text{old}} \sqrt{\mathbb{d}_j^{\text{new}} / \mathbb{d}_j^{\text{old}}}$. This is equivalent to reconstructing $(\Sigma^{\text{new}}, \alpha^{\text{new}}, \mathbf{c}^{\text{new}})$ from $(\mathbb{D}^{\text{new}}, \mathsf{R}, \mathring{\alpha}, \mathring{\mathbf{c}})$ and then imputing $\mathbf{y}_{i\text{m}} \sim f(\mathbf{y}_{i\text{m}} | \Sigma^{\text{new}}, \alpha^{\text{new}}, \mathbf{c}^{\text{new}}, W_o)$. This imputation strategy is advantageous for Gibbs sampling because it allows us to use the existing LDL decomposition of $\Sigma^{\text{old}}$ obtained in Step 1.

The resulting Gibbs-MDA sampling scheme is summarized in Algorithm 3.

**Algorithm 3: Gibbs-MDA for longitudinal binary or ordinal outcomes**

1.1) Draw $(\theta_j, \gamma_j)$ for $j = 1, \dots, p$ given the $MNIW(\nu_0, A, \mathbf{0}, M)$ prior and latent outcomes $Y_{\text{m}}$

1.2) For ordinal outcomes only, sample $\mathbf{c}_1, \dots, \mathbf{c}_p$ from the posterior in Equation (15)

2) For i=1,...,n, impute $\mathbf{y}_{im}^{\text{old}}$ from the truncated normal distribution in Equation (10) given $\{(\theta_j, \gamma_j, \mathbf{c}_j)_{j=1}^p\}$ using Li and Ghosh's [24] method. Create $\kappa_j \overset{iid}{\sim} \chi^2_{\nu_0}$. Update $(y_{1j}, \dots, y_{n_j j})$ as $\sqrt{\mathbb{d}_j^*}\,(y_{1j}^{\text{old}}, \dots, y_{n_j j}^{\text{old}})$ for $j = 1, \dots, p$, where $\mathbb{d}_j^{\text{new}} = a_j \mathsf{R}^{jj}/\kappa_j \sim IG(\nu_0/2, a_j \mathsf{R}^{jj}/2)$, $\Sigma^{jj} = [\gamma_j + \sum_{k=j+1}^p \gamma_k \beta_{kj}^2]$ is the $j$-th diagonal element of $\Sigma^{-1}$ and $\mathbb{d}_j^* = \mathbb{d}_j^{\text{new}}/\mathbb{d}_j = a_j \Sigma^{jj}/\kappa_j$.

*Note:* Steps 1.1 and 1.2 correspond to Step 1 of Algorithm 2.

We avoid complex matrix operations by using the following relationship

$$\mathbb{d}_j = \sum_{k=1}^{j-1} l_{jk}^2/\gamma_k + 1/\gamma_j,\ \Sigma^{jj} = \gamma_j + \sum_{k=j+1}^{p} \gamma_k \beta_{kj}^2,\ \mathsf{R}^{jj} = \mathbb{d}_j \Sigma^{jj},\ |\Sigma| = \prod_{j=1}^{p} \gamma_j^{-1},\ |\mathsf{R}| = \prod_{j=2}^{p} [\gamma_j \mathbb{d}_j]^{-1}. \quad (17)$$

### 3.6 Missing data imputation after dropout

The MDA algorithm provides a natural framework for handling missing data under MAR and MNAR dropout. For continuous outcomes, these imputation methods are described in Tang [9, 7]. The same methods can be applied in MVP to impute the latent outcomes $(y_{i,s_i+1}, \dots, y_{ip})$ following dropout. Missing outcomes are imputed after the algorithm converges. Because the response status $w_{ij}$ is unobserved after dropout, no truncation is required when imputing the post-dropout latent outcomes.

Below, we briefly review imputation under MAR and two commonly used control-based imputation methods: copy reference (CR) and J2R. Suppose $x_{iq}$ denotes the treatment status.

Under MAR, the latent outcomes are imputed sequentially at each visit following dropout according to the sequential regression model in Equation (2). In the control-based approaches, post-dropout latent outcomes for subjects in the control group are imputed in the same manner as under MAR.

In CR, conditional on their observed historical outcomes, subjects in the experimental treatment group are assumed to follow the control-group trajectory after dropout. Accordingly, $y_{ij}$ is imputed sequentially using (2), as under MAR, except that the treatment status is set to that of the control group for visits after dropout.

The J2R assumes that subjects lose all benefits from the treatment immediately after dropout. For subjects in the experimental treatment group, the post-dropout outcome can be obtained by removing the treatment effect from the MAR-based imputation

$$y_{ij}^{\text{J2R}} = y_{ij}^{\text{MAR}} - \alpha_{jq}$$

for visits after dropout, where $\alpha_{jq}$ is the posterior draw of the treatment effect from the MCMC chain.

The missing response status $w_{ij}$'s are imputed as $w_{ij} = k$ if $c_{j,k-1} < y_{ij} \le c_{jk}$ or equivalently if $\mathring{c}_{j,k-1} < \mathring{y}_{ij} \le \mathring{c}_{jk}$.

# 4 Numerical examples

## 4.1 Example 1. MMRM for continuous outcome

We analyze data from an antidepressant clinical trial. The Hamilton 17-item depression rating scale (HAMD-17) was assessed at baseline and weeks 1, 2, 4 and 6. There are 84 subjects in the experimental treatment group and 88 in the placebo group, with dropout rates of 24% (20/84) and 26% (23/88), respectively. There is only one intermittent missing value.

In MMRM, the response outcome is the change from baseline in HAMD-17 score at weeks 1, 2, 4 and 6 (i.e. $p = 4$), and the covariates include intercept, baseline HAMD-17 score and treatment status (i.e. $q = 3$). We assess the impact of uninformative and weakly informative priors on the parameter estimates and missing data imputations. Jeffrey's prior and the $IW(p + 1, I)$ prior are considered for $\Sigma$. The Jeffrey's prior is often used in missing data imputation in clinical trials [8]. The $IW(p+1, I)$ prior is chosen because the resulting marginal prior for any individual correlation is uniform [22]. The prior on the regression coefficient $\alpha$ is $\alpha|\Sigma \sim MN(\mathbf{0}, \Sigma, M^{+})$ with $M = \text{diag}(\varrho, \varrho, \varrho)$. We set $\varrho = 0, 10^{-12}, 10^{-6}, 0.01, 0.1, 0.5$. It is a flat prior with $r = \text{rank}(M) = 0$ at $\varrho = 0$, and a normal prior with $r = \text{rank}(M) = 3$ when $\varrho > 0$.

Table 3 reports the posterior mean $\pm$ standard deviation (SD) for the model parameters and imputed responses. Due to limited space, we present only $(\theta_4, \gamma_4)$ and $y_{i4}$ for the subject with an intermittent missing value. As the difference is subtle, the results are evaluated based on 1,000,000 posterior samples after a burn-in period of $20,000$ iterations, and it takes less than 2 minutes to run per scenario.

When using the same prior for $\alpha$, employing an inverse Wishart prior for the covariance matrix $\Sigma$ yields a slightly larger estimate of the precision parameter $\gamma_4$, and leads to reduced variability in the posterior samples for regression coefficients and imputed responses compared to using Jeffrey's prior for $\Sigma$. On the other hand, when we fix the prior for $\Sigma$, applying a flat prior (i.e. $M = \mathbf{0}$) on $\alpha$ results in a slightly smaller estimate of the precision parameter $\gamma_4$, and a minor increase in variability in both regression coefficients and imputed responses compared to using a normal prior on $\alpha$. The numerical result is consistent with the theoretical argument in Appendix A.3.

When a normal prior is used, the results are generally insensitive when the prior parameter $\varrho$ lies within the range of $10^{-12}$ to 0.01, with notable differences emerging at $\varrho = 0.1$ and 0.5.

Tang [26] further demonstrates through simulation that the choice of the prior may considerably affect the estimation of the precision of the treatment effect and type I error in MI with small samples.

## 4.2 Example 2. Simulated trial with longitudinal ordinal outcomes

We generate a randomized trial with $n = 300$ subjects, who are equally randomized to either the placebo ($g_i = 0$) or experimental ($g_i = 1$) treatment group. The latent variables at three post-baseline visits for each subject are assumed to follow the multivariate normal distribution

$$\begin{bmatrix} \mathring{y}_{i1} \\ \mathring{y}_{i2} \\ \mathring{y}_{i3} \end{bmatrix} \sim N\left( \begin{bmatrix} 0.1g_i \\ 0.3g_i \\ 0.6g_i \end{bmatrix}, R = \begin{bmatrix} 1 & 0.6 & 0.48 \\ 0.6 & 1 & 0.8 \\ 0.48 & 0.8 & 1 \end{bmatrix} \right) \tag{18}$$

where $\mathbf{x}_i = (1, g_i)$ and $\mathring{\alpha} = \begin{bmatrix} \mathring{\alpha}_{11} & \mathring{\alpha}_{12} \\ \mathring{\alpha}_{21} & \mathring{\alpha}_{22} \\ \mathring{\alpha}_{31} & \mathring{\alpha}_{32} \end{bmatrix} = \begin{bmatrix} 0 & 0.1 \\ 0 & 0.3 \\ 0 & 0.6 \end{bmatrix}$.

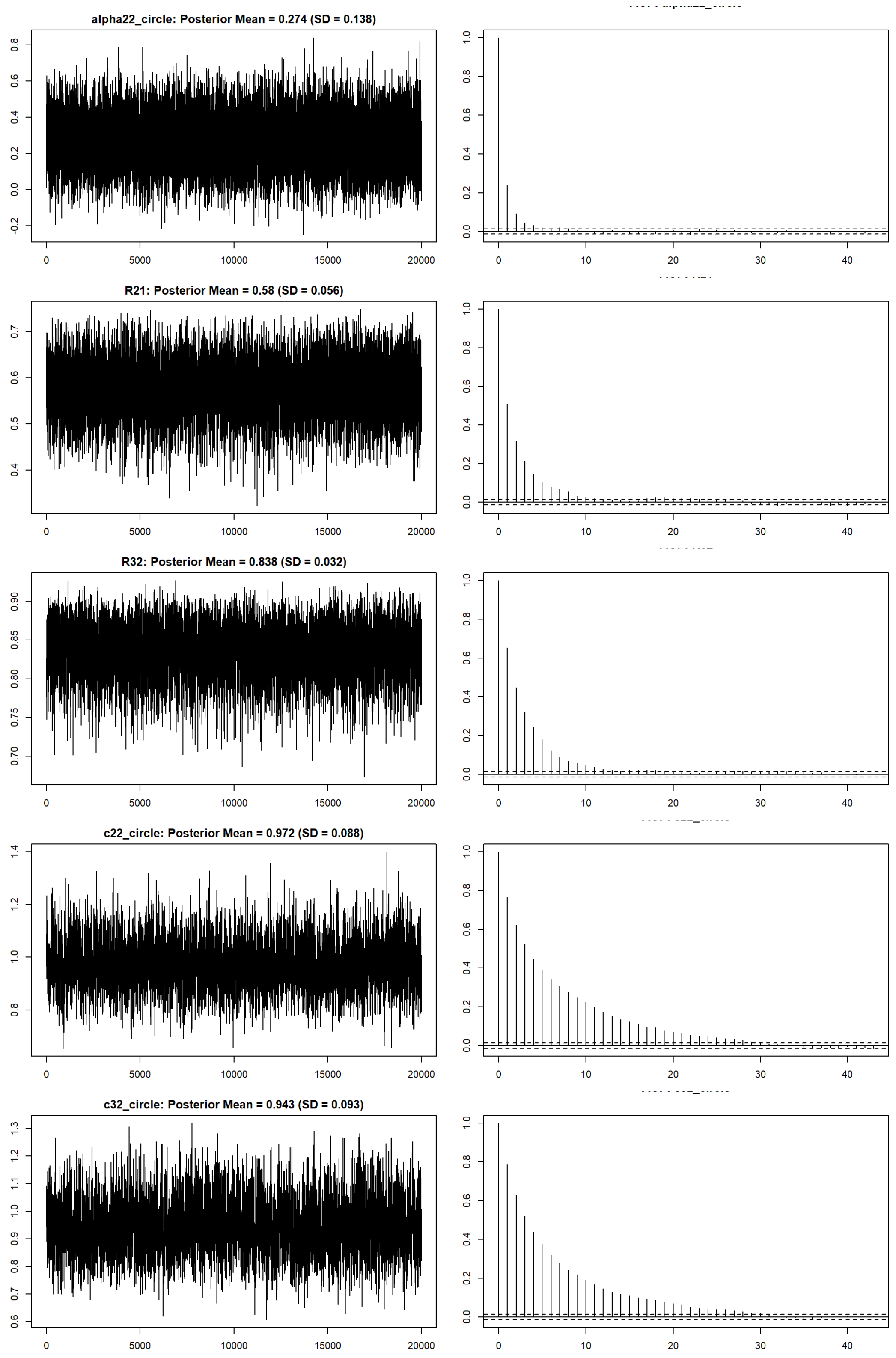

Figure 1: Trace and autocorrelation function (ACF) plots for selected parameters ($\mathring{\alpha}_{22}$, $R_{21}$, $R_{32}$, $\mathring{c}_{22}$, $\mathring{c}_{32}$) in Example 2

| $\varrho$ | $\tilde{\alpha}_{41}$ | $\tilde{\alpha}_{42}$ | $\tilde{\alpha}_{43}$ | $\beta_{41}$ | $\beta_{42}$ | $\beta_{43}$ | $\gamma_4$ | $y_{i4}$ |
|---|---|---|---|---|---|---|---|---|
| Jeffery prior on $\Sigma$: $A = \mathbf{0}$, $\nu_0 = 0$ | | | | | | | | |
| 0 | -1.973±1.184 | 0.046±0.067 | -0.977±0.706 | 0.127±0.100 | 0.170±0.086 | 0.719±0.077 | 0.070±0.009 | -0.421±3.932 |
| $10^{-12}$ | -1.973±1.170 | 0.046±0.066 | -0.977±0.698 | 0.127±0.098 | 0.170±0.085 | 0.719±0.077 | 0.071±0.009 | -0.421±3.884 |
| $10^{-6}$ | -1.973±1.170 | 0.046±0.066 | -0.977±0.698 | 0.127±0.098 | 0.170±0.085 | 0.719±0.077 | 0.071±0.009 | -0.421±3.884 |
| 0.01 | -1.971±1.169 | 0.046±0.066 | -0.977±0.698 | 0.127±0.098 | 0.170±0.085 | 0.719±0.077 | 0.071±0.009 | -0.420±3.884 |
| 0.1 | -1.955±1.164 | 0.045±0.066 | -0.975±0.697 | 0.126±0.098 | 0.170±0.085 | 0.719±0.077 | 0.071±0.009 | -0.411±3.884 |
| 0.5 | -1.886±1.143 | 0.041±0.065 | -0.967±0.692 | 0.125±0.098 | 0.170±0.085 | 0.719±0.077 | 0.071±0.009 | -0.371±3.884 |
| IW prior on $\Sigma$: $A = I(4)$, $\nu_0 = p + 1 = 5$ | | | | | | | | |
| 0 | -1.972±1.161 | 0.046±0.065 | -0.977±0.693 | 0.127±0.098 | 0.170±0.085 | 0.718±0.076 | 0.072±0.009 | -0.423±3.854 |
| $10^{-12}$ | -1.972±1.148 | 0.046±0.065 | -0.977±0.685 | 0.127±0.097 | 0.170±0.084 | 0.718±0.075 | 0.074±0.009 | -0.422±3.811 |
| $10^{-6}$ | -1.972±1.148 | 0.046±0.065 | -0.977±0.685 | 0.127±0.097 | 0.170±0.084 | 0.718±0.075 | 0.074±0.009 | -0.422±3.811 |
| 0.01 | -1.971±1.147 | 0.046±0.065 | -0.977±0.685 | 0.127±0.097 | 0.170±0.084 | 0.718±0.075 | 0.074±0.009 | -0.421±3.811 |
| 0.1 | -1.954±1.143 | 0.045±0.064 | -0.975±0.684 | 0.126±0.097 | 0.170±0.084 | 0.718±0.075 | 0.074±0.009 | -0.411±3.811 |
| 0.5 | -1.885±1.122 | 0.041±0.063 | -0.967±0.679 | 0.125±0.097 | 0.171±0.084 | 0.719±0.075 | 0.074±0.009 | -0.371±3.811 |

Table 3: Posterior mean ± standard deviation of parameter estimates and imputed values under non-informative and weakly informative Priors in MMRM

The observed ordinal response $w_{ij}$ is defined by

$$w_{ij} = \begin{cases} 1, & \text{if } \mathring{y}_{ij} \leq 0, \\ 2, & \text{if } 0 < \mathring{y}_{ij} \leq 1, \\ 3, & \text{if } \mathring{y}_{ij} > 1. \end{cases} \tag{19}$$

The cut points for the resoponse are $\mathring{c}_{12} = \mathring{c}_{22} = \mathring{c}_{32} = 1$.

Monotone missing data are generated according to the following mechanism.

$$\Pr(w_{ij} \text{ is missing}) = \frac{1}{1 + \exp(-1.5 - 0.3 w_{i,j-1})} \text{ for } j = 2, 3 \tag{20}$$

This trial includes 30, 34, and 264 subjects with one, two, and three observed visits, respectively.

The trial is analyzed using the MVP with the Gibbs-MDA Algorithm 3. We set $M = \text{diag}(\varrho, \varrho)$ with $\varrho = 0.001$, and assign the prior $c_{j2} \sim N(0, 100^2)$ for $j = 1, 2, 3$. We obtain $20,000$ posterior samples following a burn-in period of $20,000$ iterations. Figure 1 displays the trace and autocorrelation function (ACF) plots for 5 selected parameters when $c_{j2}$ is is sampled from the posterior distribution in Equation (15). Similar results (not shown) are obtained when $c_{j2}$ is drawn from the approximate uniform posterior distribution in Equation (16). These plots indicate good convergence and mixing, with the autocorrelations decaying to below 0.1 by lag 20. Furthermore, the posterior means of all parameters are fairly close to their corresponding true values, as reported in the trace plot titles.

We repeat the simulation study independently 100 times. For each dataset, we compute the posterior means of the model parameters via MCMC sampling. Across the 100 replicates, the empirical means ± SD for the subset of parameters shown in Figure 1 are $\mathring{\alpha}_{22} = 0.267 \pm 0.134$, $R_{21} = 0.593 \pm 0.048$, $R_{32} = 0.797 \pm 0.038$, $\mathring{c}_{22} = 1.014 \pm 0.090$ and $\mathring{c}_{32} = 1.004 \pm 0.104$. These estimates are close to their true values, indicating that the proposed method yields reliable parameter estimates.

### 4.3 Example 3. MVP for longitudinal binary outcomes

Lu and Chen [17] compare five MCMC algorithms for MVP using the antidepressant trial data in Example 1. The binary endpoint for HAMD-17 remission at each visit is defined as achieving a total score less than or equal to 7. Following their approach, we assume the remission statuses $w_{ij}$'s are collected, but the HAMD-17 scores $y_{ij}$'s are missing at week 1, 2, 3 and 6. The baseline covariate vector is $\mathbf{x}_i = (1, \text{baseline HAMD-17 score}, \text{treatment status})$. In certain clinical scenarios, the endpoint of interest is the binary outcome $w_{ij}$ while the continuous outcomes $y_{ij}$ for defining $w_{ij}$ are collected. In such cases, it is recommended to perform multiple imputations based on the continuous outcomes $y_{ij}$ using MMRM rather than based on the categorical endpoints $w_{ij}$ using MVP.

We compare the MDA algorithm with the FDA-based parameter-expanded Gibbs (f-PX-Gibbs) algorithm, originally developed by Talhouk et al. [25] and subsequently studied by Lu and Chen [17]. The prior parameter is $M = \text{diag}(\varrho, \varrho, \varrho)$ with $\varrho = 0.01$. The result changes little when we set $\varrho = 0.001$.

The MCMC schemes implemented in Lu and Chen [17] are not efficient. At each MCMC iteration, Lu and Chen [17] initialize $y_{ij}$ as $-1$, 0 or 1, and use a Gibbs subchain with $l_y = 5$ or 50 cycles to update $(y_{i1}, \dots, y_{ip})$. This necessitates achieving the target distribution for $(y_{i1}, \dots, y_{ip})$ at every MCMC iteration, which is generally difficult to guarantee in practice. A large $l_y$ is particularly needed when clinical outcomes are collected at many visits, posing difficulty in choosing $l_y$ in advance and inducing inefficiency. The f-PX-Gibbs algorithm (a bug on calculating the scale matrix for sampling $\Sigma$ in Lu and Chen's SAS code is fixed) yields parameter estimates close to the MDA estimates at $l_y = 50$, but not at $l_y = 30$. Modifying their code to use the value of $(y_{i1}, \dots, y_{ip})$ from the previous MCMC iteration as the initial value in the current iteration (with $l_y = 1$ or 2) allows the FDA scheme to yield very similar results to the MDA scheme. This simple adjustment also reduces the f-PX-Gibbs runtime from up to one hour to about two minutes.

The MDA and FDA algorithms have the same stationary distribution. We do not expect systematic differences in inference between the two algorithms. However, the two algorithms differ in sampling efficiency. Figure 2 shows the ACF plots for two representative parameters, based on $20,000$ consecutive posterior samples collected after a burn-in period of $20,000$ iterations. The ACFs decay more rapidly under MDA than under FDA, indicating better mixing of the MDA chains. The ACFs from the revised FDA algorithm appear to decay more rapidly than those from the FDA implementation of Lu and Chen [17], as reported in their online supplementary material, possibly due to issues with the latent variable generation in Lu and Chen [17] discussed above. The trace plots also indicate satisfactory convergence of both MCMC chains, which are omitted due to space limitations.

Table 4 displays the posterior mean of the parameter estimates from MDA (the posterior SD is omitted due to limited space) and the MI estimate of the difference in the remission rate at week 6 under MAR, CR and J2R [17, 9]. The results are based on $10,000$ posterior samples obtained by retaining every 10th iteration after a burn-in period of $50,000$ iterations. The standard errors of the treatment effect estimates are fairly similar across the three MI analyses. The estimated difference in remission rates is substantially lower under J2R than under MAR. In this example, the CR imputation does not yield more conservative estimates than the MAR imputation. It is because the posterior mean of $\tilde{\alpha}_{43} = -0.091 < 0$, indicating that, conditional on the history $(\mathbf{x}_i, y_{i1}, y_{i2}, y_{i3})$, subjects in the experimental arm are not expected to gain additional benefit relative to those in the placebo arm under MAR.

Lu and Chen [17] consider the Bayesian estimate of the rate difference under MAR, defined as

$$\Delta_{MAR} = n^{-1} \sum_{i=1}^{n} \Phi\left( \sum_{a=1}^{q-1} x_{ia} \mathring{\alpha}_{pa} + \mathring{\alpha}_{pq} \right) - n^{-1} \sum_{i=1}^{n} \Phi\left( \sum_{a=1}^{q-1} x_{ia} \mathring{\alpha}_{pa} \right)$$

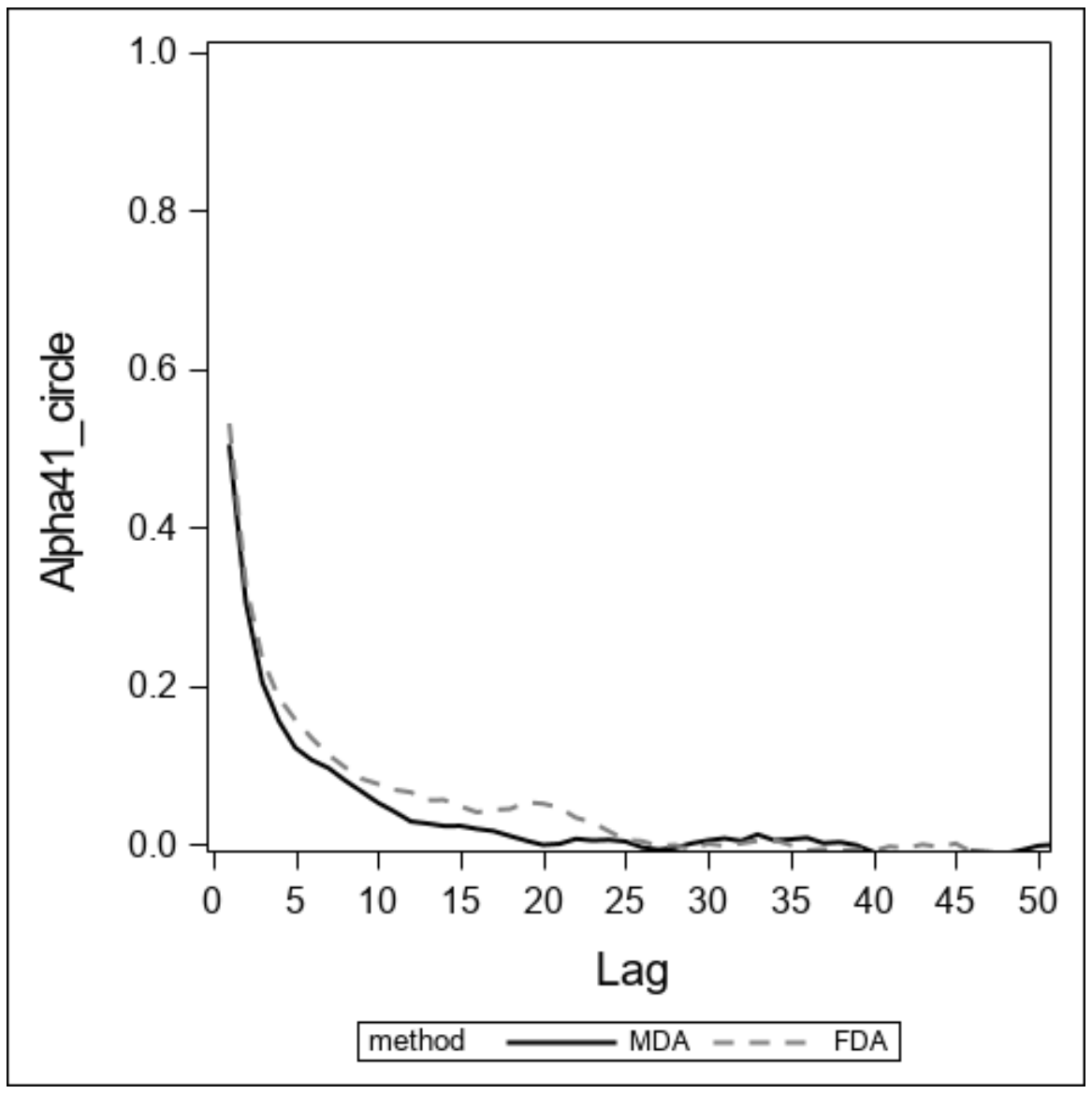

(a) $\mathring{\alpha}_{41}$

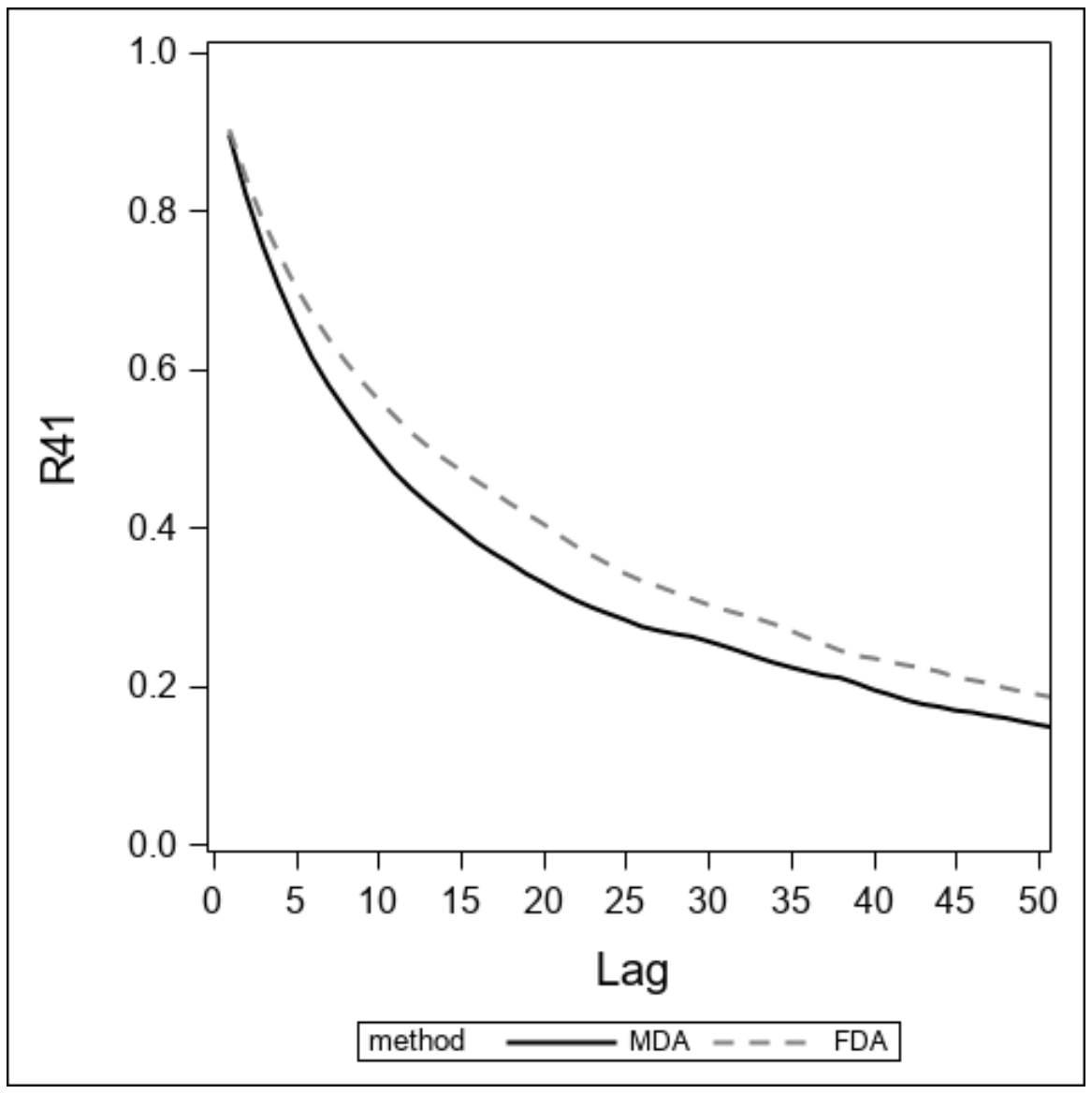

(b) $r_{41}$

Figure 2: Autocorrelation for two parameters from the MDA and FDA algorithms

where $x_{iq} = 1$ or 0 represents treatment status. The Bayes estimate $\pm$ SD under MAR by Lu and Chen [17] is $0.061 \pm 0.071$, showing notable differences from the MI estimate. In the analysis of confirmatory clinical trials, the frequentist MI approach is commonly employed.

This example demonstrates that the MDA and FDA schemes yield comparable inference results because they target the same posterior distribution, while MDA achieves greater sampling efficiency. It also shows that J2R can be fairly conservative relative to MAR, as treatment benefits are assumed to be lost after dropout under J2R, but maintained under MAR.

# 5 MH algorithms under a general prior

We derive the Gibbs-MDA algorithm in MVP under a specific yet broadly applicable prior in Section 3. For a more general prior, the posterior sampling can be achieved via the iMH sampler. For example, we resort to the iMH sampler if normal priors are placed on $\mathring{\alpha}$ with non-zero prior means. Note that the f-PX-Gibbs algorithm of Lu and Chen [17] for longitudinal binary outcomes is incorrect when $\mathring{\alpha}_0 \neq \mathbf{0}$.

We consider a general prior for $(\mathsf{R}, \mathring{\alpha})$

$$p(\mathsf{R}, \mathring{\alpha}) \propto g(\mathsf{R}, \mathring{\alpha}) \left\{ |\mathsf{R}|^{-[\nu_0+p+1]/2} \prod_{j=1}^{p} \left[\mathsf{R}^{jj}\right]^{-\nu_0/2} \right\} \left\{ |\mathsf{R}|^{-q/2} \exp[-\mathrm{tr}\{M(\mathring{\alpha} - \mathring{\alpha}_0)' \mathsf{R}^{-1} (\mathring{\alpha} - \mathring{\alpha}_0)\}] \right\} \qquad (21)$$

where $M$ is a full-rank matrix. When $g(\mathsf{R}, \mathring{\alpha}) = 1$, the prior is similar to that defined in Equation (11) except that the prior mean $\mathring{\alpha}_0$ for $\mathring{\alpha}$ may not be zero.

Similar to the derivation of Equation (14), we obtain the following posterior distribution,

$$\pi(\Sigma, \alpha | Y_{\mathrm{m}}, W_o) \propto \varphi(\mathsf{R}, \mathring{\alpha}, \mathbb{D})\, \psi(\Sigma, \alpha | \nu_0, A, \alpha_0^*, M) \qquad (22)$$

| Posterior mean for sequential regression of latent $y_{ij}$ on $\mathbf{x}_i$, $y_{i1},\dots,y_{i,j-1}$ | | | | | | | | |
|---|---|---|---|---|---|---|---|---|
| $j$ | $\tilde{\alpha}_{j1}$ | $\tilde{\alpha}_{j2}$ | $\tilde{\alpha}_{j3}$ | $\beta_{j1}$ | $\beta_{j2}$ | $\beta_{j3}$ | $\gamma_j$ | $d_j$ |
| 1 | 0.831 | -0.188 | 0.281 | . | . | . | 1.122 | 1.672 |
| 2 | 1.220 | -0.020 | 0.106 | 1.440 | . | . | 1.746 | 1.656 |
| 3 | -0.859 | 0.069 | 0.147 | -0.251 | 1.157 | . | 2.649 | 1.687 |
| 4 | -0.088 | -0.003 | -0.091 | -0.105 | 0.176 | 0.738 | 5.002 | 1.613 |

| Posterior mean of model parameters from MMRM regression of latent $(\mathring{y}_{i1},\dots,\mathring{y}_{i4})$ on $\mathbf{x}_i$ | | | | | | | |
|---|---|---|---|---|---|---|---|
| $j$ | $\mathring{\alpha}_{j1}$ | $\mathring{\alpha}_{j2}$ | $\mathring{\alpha}_{j3}$ | $R_{j1}$ | $R_{j2}$ | $R_{j3}$ | $R_{j4}$ |
| 1 | 0.643 | -0.144 | 0.209 | 1.000 | 0.824 | 0.680 | 0.632 |
| 2 | 1.104 | -0.128 | 0.235 | 0.824 | 1.000 | 0.875 | 0.826 |
| 3 | 0.645 | -0.075 | 0.275 | 0.680 | 0.875 | 1.000 | 0.910 |
| 4 | 0.670 | -0.077 | 0.197 | 0.632 | 0.826 | 0.910 | 1.000 |

| MI estimate of difference in remission proportion $\pm$ standard error at week 6 under MAR and MNAR | | |
|---|---|---|
| MAR | Copy Reference | Jump to Reference |
| $0.029 \pm 0.075$ | $0.029 \pm 0.074$ | $0.017 \pm 0.073$ |

Table 4: Estimated model parameters in MVP, and MI analysis under MAR and control-based imputation

where $\alpha_0^*$ is a $p \times q$ matrix of fixed values, and

$$\begin{aligned} &\psi(\Sigma, \alpha | \nu_0, A, \alpha_0^*, M) \propto MNIW(\alpha, \Sigma | \nu_0, A, \alpha_0^*, M) \prod_{i=1}^{n} N(\mathbf{y}_{i\mathrm{m}} | \alpha_i \mathbf{x}_i', \Sigma_i). \\ &\varphi(\mathsf{R}, \mathring{\alpha}, \mathbb{D}) = g(\mathsf{R}, \mathring{\alpha}) \exp(\mathrm{tr}[M(\alpha - \alpha_0^*)' \Sigma^{-1} (\alpha - \alpha_0^*) - M(\alpha - \mathbb{D}\mathring{\alpha}_0)' \Sigma^{-1} (\alpha - \mathbb{D}\mathring{\alpha}_0)]/2). \end{aligned} \tag{23}$$

We choose $A = I$. As discussed in Appendix A.4, additional terms also arise in $\varphi(\mathsf{R}, \mathring{\alpha}, \mathbb{D})$ when the prior is specified on $\mathring{\mathbf{c}}_j$ rather than on $\mathbf{c}_j$.

We use the iMH sampler [27] to draw $(\Sigma, \alpha)$. We generate the candidate $(\alpha, \Sigma)$ from the proposal distribution $\psi(\alpha, \Sigma)$. We update $(\Sigma, \alpha)$ as $(\Sigma^{\mathrm{cand}}, \alpha^{\mathrm{cand}})$ with a probability determined by the $\varphi(\mathsf{R}, \mathring{\alpha}, \mathbb{D})$ function, or retain it unchanged otherwise. In the proposal distribution, we set $\alpha_0^* = \mathbf{0}$ for a noninformative or weakly informative prior on $\alpha$. If the prior for $\alpha$ is informative, and the acceptance rate is low, we may modify $\alpha_0^*$ after a certain number of MCMC iterations, and subsequently fix the value of $\alpha_0^*$. The iMH scheme is presented in Algorithm 4.

**Algorithm 4: iMH-MDA for longitudinal binary or ordinal outcomes**

1.1) Sample $(\alpha, \Sigma)$ by iMH sampler: Draw candidate $(\theta_j^{\text{cand}}, \gamma_j^{\text{cand}})$ from the MDA algorithm for MMRM under the $MNIW(\nu_0, A, \alpha_0^*, M)$ prior. Set $(\theta_j, \gamma_j)$ to $(\theta_j^{\text{cand}}, \gamma_j^{\text{cand}})$ with probability $\min\left\{\frac{\varphi(\mathsf{R}^{\text{cand}}, \mathring{\alpha}^{\text{cand}}, \mathbb{D}^{\text{cand}})}{\varphi(\mathsf{R}^{\text{curr}}, \mathring{\alpha}^{\text{curr}}, \mathbb{D}^{\text{curr}})}, 1\right\}$. Otherwise, keep $(\theta_j, \gamma_j)$ unchanged.

1.2) For ordinal outcomes only, sample $\mathbf{c}_1, \ldots, \mathbf{c}_p$ from the posterior in Equation (15)

2) Rescale $(\theta_j, \gamma_j, \mathbf{c}_j)$'s corresponds to the change from $(\mathbb{d}_1, \ldots, \mathbb{d}_p)$ to $(\mathbb{d}_1^{\text{new}}, \ldots, \mathbb{d}_p^{\text{new}})$, while holding $(\mathsf{R}, \mathring{\alpha}, \mathring{\mathbf{c}})$ fixed, using

$$(\tilde{\alpha}_{jk}, \beta_{jk}, c_{jk}, l_{jk}, \gamma_j) \rightarrow (\sqrt{\mathbb{d}_j^*}\tilde{\alpha}_{jk}, \sqrt{\mathbb{d}_j^*/\mathbb{d}_k^*}\beta_{jk}, \sqrt{\mathbb{d}_j^*}c_{jk}, \sqrt{\mathbb{d}_j^*/\mathbb{d}_k^*}l_{jk}, \gamma_j/\mathbb{d}_j^*) \quad (24)$$

where $\mathbb{d}_j^* = \mathbb{d}_j^{\text{new}}/\mathbb{d}_j$ is defined in Step 2 of Algorithm 3, and $l_{jk}$ is the $(l, k)$-th element of $L$. Impute $\mathbf{y}_{im}$ from Equation (10) given the updated $(\theta_j, \gamma_j, \mathbf{c}_j)$'s by Li and Ghosh's method.

*Note:* [1] In Step 1.1, if $\varphi(\mathsf{R}, \mathring{\alpha}, \mathbb{D}) = g(\mathsf{R})$ is solely a function of $\mathsf{R}$, we can draw $(\beta_j, \gamma_j)$ marginally via the iMH algorithm and then sample $\tilde{\alpha}_j$ conditioning on $(\beta_j, \gamma_j)$. [2] In Step 2, imputation of $\mathbf{y}_{im}$ is equivalent in Algorithms 3 and 4. However, $(\Sigma, \alpha)$ is additionally updated for MH sampling in Algorithm 4.

In the MDA algorithm, drawing $(\Sigma, \alpha)$ is achieved through drawing $(\theta_j, \gamma_j)$'s. The acceptance probability function is invariant under reparameterization since we apply the same Jacobian of the transformation $(\alpha, \Sigma) \rightarrow (\theta_j, \gamma_j)$'s to both the target and proposal distributions. It is possible to avoid reconstructing $(\alpha, \Sigma)$ by using the the analytic formula in Equation (17).

In step (1), we have the option to sample $(\theta_j, \gamma_j)$ either simultaneously for all $j$ or sequentially for $j = 1, \ldots, p$. The same $\varphi(\cdot)$ function is employed to calculate the acceptance probability except that $\{(\theta_j, \gamma_j) : j \neq l\}$ is held fixed at their current values during the update of $(\theta_l, \gamma_l)$. When the prior is only weakly informative, the proposal distribution is generally close to the target distribution. In our limited experience, the simultaneous approach typically achieves an acceptance rate above 85%, with ACFs that are not obviously inferior to those from the Gibbs sampler.

When $\mathring{\alpha}_0 = \mathbf{0}$ and $g(\mathsf{R}, \mathring{\alpha})$ is solely a function of $\mathsf{R}$, then $\alpha_0^* = \mathbf{0}$ and $\varphi(\mathsf{R}, \mathring{\alpha}, \mathbb{D})$ simplifies to $g(\mathsf{R})$. We can draw $(\beta_j, \gamma_j)$'s marginally using the iMH sampler, and then sample $\tilde{\alpha}_j$'s conditioning on $(\beta_j, \gamma_j)$ by the Gibbs sampler. The candidate $(\beta_j, \gamma_j)$ can be sampled using the method in Appendix A.1, and the acceptance probability function is $\min\left\{\frac{g(\mathsf{R}^{\text{cand}})}{g(\mathsf{R}^{\text{curr}})}, 1\right\}$. Again, these updates can be performed simultaneously for all $j$ or sequentially for $j = 1, \ldots, p$.

# 6 Discussion

We extend Tang's [6, 7] MDA algorithm for longitudinal continuous outcomes to longitudinal binary and ordinal endpoints. In MVP, the longitudinal categorical outcomes are discretized versions of underlying latent Gaussian outcomes modeled by MMRM. Once the latent variables $Y_{\text{m}}$ prior to dropout are imputed, the MVP model reduces to an MMRM framework. Under a specific yet broadly applicable prior defined in Equation (3.3), $(\Sigma, \alpha)$ or equivalently $(\theta_j, \gamma_j)$'s in MVP can be drawn by the same sampler for longitudinal continuous outcomes. This property aids in understanding the algorithm and developing computer codes for MVP. Furthermore, the MAR and MNAR imputation methods developed for continuous endpoints [9, 7] can be directly applied to impute the post-dropout latent variables. As demonstrated in our examples, the proposed method provides a practical framework for Bayesian analysis and missing data imputation, accommodating both MAR and MNAR assumptions commonly considered in clinical trials.

To accommodate complex priors on $(\alpha, \Sigma)$, we use the iMH sampler with candidate values generated from the Gibbs sampler. The iMH sampler can similarly be used for the cutoff parameters $\mathring{\mathbf{c}}_j$ when their priors are complex. However, it is not appropriate to first generate $(\alpha, \Sigma)$ marginally and then use the iMH sampler to draw $\mathbf{c}_j$ conditional on $(\alpha, \Sigma)$ [28]. Instead, $\mathbf{c}_j$'s can be sampled jointly with $(\alpha, \Sigma)$ using the iMH sampler or conditionally on $(\alpha, \Sigma)$ using the iterated iMH approach [28].

The MDA scheme yields parameter estimates that closely match those produced by our revised version of the f-PX-Gibbs scheme in the numerical analysis. This agreement is expected, as Bayesian inference is determined entirely by the posterior distribution. Lu and Chen [17] report discrepancies across different MCMC algorithms. The discrepancies arise partly from differences in the priors and partly because the posterior distribution derived by Lu and Chen for the MDA algorithm is incorrect when non-flat priors are placed on the regression coefficients. In our previous work [13], we inadvertently omitted these prior terms in deriving the posterior distribution. This omission was subsequently identified and corrected. See also Section 4.3 for additional implementation issues in Lu and Chen [17].

The superiority of the MDA scheme over the FDA scheme is supported by the theoretical results of Liu [11] and Liu et al. [12], as well as by our numerical analysis. The MDA scheme also incurs a lower computational cost per MCMC iteration. Lu and Chen [17], however, report that their FDA-based scheme is more efficient than the MDA algorithm of Tang [13], possibly because their implementation uses SAS IML, where computational speed depends not only on the amount of computation but also on the number of matrix operations and function calls. For example, $(\Sigma, \alpha)$ can be sampled directly using built-in SAS IML functions in f-PX-Gibbs, substantially reducing matrix operations. We optimized our MDA implementation to slightly outperform the (revised) FDA scheme by introducing a few redundant computations when imputing latent variables, thereby reducing the total number of matrix operations and computation time. Compared to our original MDA scheme [13], the new MDA scheme has a simpler and more easily interpretable posterior distribution, and is therefore preferred.

In the proposed MCMC algorithm, several strategies are employed to improve sampling efficiency. First, the PX scheme allows the Markov chain to move more freely in an enlarged parameter space, thereby improving mixing. It also guarantees that the correlation matrix $\mathsf{R}$ is positive definite. Second, $(\Sigma, \alpha, \mathbf{c}_1, \dots, \mathbf{c}_p)$ are jointly updated as a block. Third, we integrate out the latent variables after dropout within the MDA framework. Fourth, the Li–Ghosh approach reparameterizes the latent vector $(y_{i1}, \dots, y_{is_i})$ to decorrelate the parameter space and improve sampling efficiency. The conventional sampling scheme for the cut points $\mathring{\mathbf{c}}_j$ can suffer from high autocorrelation due to the narrow support [29, 30] of their posterior distributions (see Equation (16)). Simulation 2 indicates that the PX scheme effectively reduces autocorrelation in $\mathring{\mathbf{c}}_j$. Cowles and Carlin [29] and Lee et al. [30] use a hybrid Gibbs/MH algorithm to jointly sample $(\mathring{\mathbf{c}}_j, \mathring{Y})$ to improve the mixing of $\mathring{\mathbf{c}}_j$.

For MMRM, we show that diffuse normal priors for the regression coefficients $\alpha$ tend to yield regression coefficient estimates and imputed responses with smaller posterior variability than flat priors. In contrast, flat priors can lead to numerical instability in the presence of multicollinearity. For MVP, we recommend normal priors over flat or highly diffuse priors, as the latter may result in an improper posterior that is difficult to detect in practice [31].

For ordinal outcomes, we allow the cut points $\mathring{\mathbf{c}}_j$ to differ across visits. This time-varying specification is motivated by the observation in clinical trials that both the distribution and variability of continuous outcomes often change over time as disease progresses. This flexible parameterization is well-suited for large trials. For small samples, enforcing common cutoffs across visits can stabilize estimation; however, it introduces constraints on the expansion parameters $\mathbb{d}_j$:

$$\max_{\{(i,j):\, w_{ij}=k\}} \frac{y_{ij}}{\mathbb{d}_j} \leq \min_{\{(i,j):\, w_{ij}=k+1\}} \frac{y_{ij}}{\mathbb{d}_j}, \qquad k = 1, \dots, K-1.$$

Because these constraints reduce the efficiency of the PX algorithm, developing alternative MCMC schemes that bypass them is an important direction for future research. Other directions include extending the algorithm to accommodate missing covariates, and conducting a systematic comparison with existing MAR and MNAR imputation approaches for longitudinal categorical outcomes.

# A Technical Details

## A.1 Random Number Generation from the NG Distribution

We discuss the generation of NG random variables based on Tang [6]. In practice, we know $D$. There is no need to generate $\gamma$ first, and then draw $\theta$ given $\gamma$.

Let $D = BB'$ be the Cholesky decomposition of $D$, $\begin{bmatrix} B_{\theta\theta} & 0 \\ B_{\gamma\theta} & B_{\gamma\gamma} \end{bmatrix}$ a partition of $B$ according to $\begin{bmatrix} \theta \\ \gamma \end{bmatrix}$, and $a = B_{\gamma\gamma}^2$. To draw random NG variables from the distribution (5), we create $e_j \overset{iid}{\sim} N(0,1)$ for $j \leq m-1$ and $e_m^2 \sim \chi_f^2$. Let $\mathbf{e}_\theta = (e_1, \dots, e_m)'$ and $\mathbf{e}_\beta = (e_{q+1}, \dots, e_m)'$.

We can generate $(\gamma, \theta)$ as

$$(h_1, \dots, h_m)' = (B')^{-1}\mathbf{e}_\theta, \ \gamma = h_m^2 \text{ and } \theta = -(h_1, \dots, h_{m-1})'/h_m.$$

Alternatively, we can generate $(\gamma, \beta)$ marginally and then $\tilde{\alpha}$ conditioning on $(\gamma, \beta)$ as

$$(h_{q+1}, \dots, h_m)' = (B_{22}')^{-1}\mathbf{e}_\beta, \ \gamma = h_m^2, \ \beta = -(h_{q+1}, \dots, h_{m-1})'/h_m \text{ and } \tilde{\alpha} = (B_{\alpha\alpha}')^{-1}[\mathbf{e}_\alpha/\sqrt{\gamma} + B_{21}'\beta]$$

where $\begin{bmatrix} B_{\alpha\alpha} & \mathbf{0} \\ B_{21} & B_{22} \end{bmatrix}$ is a partition of $B$ with $B_{\alpha\alpha}$ having dimensions $q \times q$, and $\mathbf{e}_\alpha = (e_1, \dots, e_q)'$.

We use the first approach in the Gibbs sampler. The second approach can nevertheless be useful for certain complex priors. As discussed in Section 5, one can sample $\Sigma$, or equivalently the $(\beta_j, \gamma_j)$'s, marginally using the iMH sampler and then draw $\tilde{\alpha}_j$ conditional on $\Sigma$ using the Gibbs sampler.

## A.2 FDA scheme for longitudinal continuous outcomes

**Lemma A** *In MMRM, the posterior distribution of* $(\Sigma, \alpha)$ *under the MNIW prior in Equation* (6) *given the complete data is*

$$\Sigma|Y \sim IW[n + \nu_0 + r - q, A_{pos}] \text{ and } \alpha|\Sigma, Y \sim MN(\alpha_{pos}, \Sigma, \Omega^{-1}) \tag{A1}$$

*where* $\Omega = X'X + M$, $\alpha_{pos} = (Y'X + \alpha_0 M)\Omega^{-1}$ *and* $A_{pos} = A + Y'Y + \alpha_0 M \alpha_0' - \alpha_{pos}\Omega\alpha_{pos}'$.

*For multivariate normal data, the posterior distribution simplifies to*

$$\Sigma|Y \sim IW[n + \nu_0 + r - 1, A_{pos}] \text{ and } \mu|\Sigma, Y \sim N(\mu_{pos}, (n + m_0)^{-1}\Sigma) \tag{A2}$$

*where* $\bar{\mathbf{y}} = n^{-1}\sum_{i=1}^n \mathbf{y}_i$, $\mu_{pos} = \frac{n\bar{\mathbf{y}} + m_0\mu_0}{n + m_0}$ *and* $A_{pos} = A + \sum_{i=1}^n (\mathbf{y}_i - \bar{\mathbf{y}})^{\otimes 2} + \frac{m_0 n}{m_0 + n}(\mu_0 - \bar{\mathbf{y}})^{\otimes 2}$.

The proof follows from straightforward algebra and is omitted. Note that $A_{\text{pos}}$ can be written as $A + (Y - X\hat{\alpha}')'(Y - X\hat{\alpha}') + (\alpha_0 - \hat{\alpha})[M - M\Omega^{-1}M](\alpha_0 - \hat{\alpha})'$ with $\hat{\alpha} = Y'X(X'X)^{-1}$.

**Algorithm 2: FDA for longitudinal continuous outcomes**

1 For $i = 1, \ldots, n$, jointly impute intermittent missing data $\mathbf{y}_{i\text{i}}$ and post-dropout missing data $\mathbf{y}_{i\text{d}}$ from the conditional normal distribution $f(\mathbf{y}_{i\text{i}}, \mathbf{y}_{i\text{d}} | \mathbf{y}_{io}, \alpha, \Sigma)$

2. Update $(\alpha, \Sigma)$: One classic way is to draw $\Sigma \sim \pi(\Sigma|Y)$, and then $\alpha \sim \pi(\alpha \mid \Sigma, Y)$ according to Lemma A. Alternatively, we sample $(\theta_j, \gamma_j)$ for $j = 1, \ldots, p$ from the posterior in Equation (8) given the augmented full data ($n_1 = \ldots = n_p = n$).

### A.3 Comparison of flat or diffuse normal priors in MMRM

We compare the posterior distributions in MMRM under two uninformative priors on $\alpha$ when $A = \text{diag}(a, \ldots, a)$, $\alpha_0 = \mathbf{0}$, and the data pattern is monotone. The following results can be readily derived from Lemma 1.

- Suppose we put a flat prior (i.e. $M = \mathbf{0}$) on $\alpha$. Let $M^{\text{uni}} = \text{diag}(0, \ldots, 0, a, \ldots, a)$ with the first $q$ diagonal elements given by 0 and the remaining $j - 1$ elements given by $a$, $Y_j = (y_{j1}, \ldots, y_{n_j,j})'$ and $\hat{S}_j^{\text{uni}} = Y_j'[I - Z_{j-1}(Z_{j-1}'Z_{j-1} + M_j^{\text{uni}})^{-1}Z_{j-1}']Y_j + a$. The prior for $(\gamma_j, \theta_j)$ is

$$p(\gamma_j, \theta_j) \propto \left\{\gamma_j^{(\nu_0+j-p-q)/2-1} \exp(-\gamma_j a/2)\right\} \left\{\gamma_j^{(q+j-1)/2} \exp(-\gamma_j \theta' M_j^{\text{uni}} \theta/2)\right\}$$

  The posterior distribution of $(\gamma_j, \theta_j)$ is given by

$$\gamma_j | Y_{\text{m}} \sim \chi^2_{n_j+\nu_0+j-p-q} / \hat{S}_j^{\text{uni}} \text{ and } \theta_j | \gamma_j, Y_{\text{m}} \sim N[(Z_{j-1}'Z_{j-1} + M_j^{\text{uni}})^{-1}Z_{j-1}'Y_j, \gamma_j^{-1}(Z_{j-1}'Z_{j-1} + M_j^{\text{uni}})^{-1}]$$

- Consider a diffuse normal prior on $\alpha$ with $M = \text{diag}(b, \ldots, b)$. It induces the prior for $(\gamma_j, \theta_j)$

$$p(\gamma_j, \theta_j) \propto \left\{\gamma_j^{(\nu_0+j-p)/2-1} \exp(-\gamma_j a/2)\right\} \left\{\gamma_j^{(q+j-1)/2} \exp(-\gamma_j \theta' M_j^{\text{dif}} \theta/2)\right\}$$

  where $M_j^{\text{dif}} = \text{diag}(b, \ldots, b, a, \ldots, a)$ with the first $q$ diagonal elements given by $b$ and the remaining $j - 1$ elements given by $a$, Let $\hat{S}_j^{\text{dif}} = Y_j'[I - Z_{j-1}(Z_{j-1}'Z_{j-1} + M_j^{\text{dif}})^{-1}Z_{j-1}']Y_j + a$. The posterior distribution of $(\gamma_j, \theta_j)$ is given by

$$\gamma_j | Y_{\text{m}} \sim \chi^2_{n_j+\nu_0+j-p} / \hat{S}_j^{\text{dif}} \text{ and } \theta_j | \gamma_j, Y_{\text{m}} \sim N[(Z_{j-1}'Z_{j-1} + M_j^{\text{dif}})^{-1}Z_{j-1}'Y_j, \gamma_j^{-1}(Z_{j-1}'Z_{j-1} + M_j^{\text{dif}})^{-1}]$$

When $b$ is close to 0, we have $\hat{S}_j^{\text{uni}} \approx \hat{S}_j^{\text{dif}}$ and $M^{\text{dif}} \approx M^{\text{uni}}$, but the df in the posterior chi-squared distribution for $\gamma_j$ are different under the two priors. The diffuse normal prior tends to yield larger estimate of the precision parameter than the flat prior. This difference tends to be more pronounced when the sample size $n_j$ is small, or the number $q$ of covariates is large.

Under flat or very diffuse priors, the MDA algorithm may encounter numerical instability when multicollinearity is present since the precision matrix for $\theta_j$'s, which is $\gamma_j(Z_{j-1}'Z_{j-1} + M_j^{\text{uni}})$ or $\gamma_j(Z_{j-1}'Z_{j-1} + M_j^{\text{dif}})$, can become singular or nearly singular.

## A.4 Derivation of the posterior distribution in MVP

This section shows that the joint posterior distribution of $(\Sigma, \alpha, \mathbf{c}, Y_{\mathrm{m}})$ in the PX model is

$$\pi(\Sigma, \alpha, \mathbf{c}, Y_{\mathrm{m}}|W_o) \propto \pi(\Sigma, \alpha|Y_{\mathrm{m}}, W_o) \left\{ \prod_{j=1}^{p} \pi(\mathbf{c}_j|\Sigma, \alpha, Y_{\mathrm{m}}, W_o) \right\} \tag{A3}$$

for $\pi(\Sigma, \alpha|Y_{\mathrm{m}}, W_o)$ and $\pi(\mathbf{c}_j|\Sigma, \alpha, Y_{\mathrm{m}}, W_o)$ defined respectively in Equations (14) and (15). Note that $A = \mathrm{diag}(a_1, \dots, a_p)$, with default values $a_j = 1$ and $A = I$.

The posterior distribution in the PX model is given by

$$\pi(\Sigma, \alpha, \mathbf{c}, Y_{\mathrm{m}}|W_o) \propto (J_1 J_2) \left\{ p(\mathsf{R}) p(\mathring{\alpha}|\mathsf{R}) \prod_{j=1}^{p} \left[ p(\mathring{\mathbf{c}}_j|\mathbb{d}_j, \mathsf{R}) p(\mathbb{d}_j|\mathsf{R}) \right] \prod_{i=1}^{n} f(\mathring{\mathbf{y}}_{i\mathrm{m}}|\mathsf{R}_i, \mathring{\alpha}, \mathring{\mathbf{c}}, s_i) \right\} \tag{A4}$$

where $p(\mathsf{R}) \propto |\mathsf{R}|^{-[\nu_0+p+1]/2} \prod_{j=1}^{p} [\mathsf{R}^{jj}]^{-\nu_0/2}$, $p(\mathring{\alpha}|\mathsf{R}) \propto |\mathsf{R}|^{-q/2} \exp[-\mathrm{tr}(M\mathring{\alpha}'\mathsf{R}^{-1}\mathring{\alpha})/2]$, $p(\mathbb{d}_j|\mathsf{R}) \propto [\mathsf{R}^{jj}]^{\nu_0/2} \mathbb{d}_j^{-(\nu_0/2+1)} \exp[-a_j \mathsf{R}^{jj}/(2\mathbb{d}_j)]$, $p(\mathring{\mathbf{c}}_j|\mathbb{d}_j, \mathsf{R}) \propto N(\mathring{\mathbf{c}}_j|\mathbb{d}_j^{-1/2}\mathbf{c}_{j0}, \mathbb{d}_j^{-1}V_{cj}) I(0 < \mathring{c}_{j2} < \dots < \mathring{c}_{j,K-1})$, $f(\mathring{\mathbf{y}}_{i\mathrm{m}}|\mathsf{R}_i, \mathring{\alpha}, \mathring{\mathbf{c}}, s_i) \propto N(\mathring{\mathbf{y}}_{i\mathrm{m}}|\mathbf{x}_i\mathring{\alpha}_i, \mathsf{R}_i) \prod_{j=1}^{s_i} I(\mathring{y}_{ij} \in \mathring{C}_{ij})$, $J_1 = \prod_{j=1}^{p} \mathbb{d}_j^{-(n_j+q+K-2)/2}$ is the Jacobian of the transformation $(\mathring{y}_{ij}\text{'s}, \mathring{\alpha}_{jk}\text{'s}, \mathring{c}_{jk}\text{'s}) \to (y_{ij}\text{'s}, \alpha_{jk}\text{'s}, c_{jk}\text{'s})$ and $J_2 = \prod_{j=1}^{p} \mathbb{d}_j^{-(p-1)/2}$ is the Jacobian for $(\mathbb{D}, \mathsf{R}) \to \Sigma$.

Equation (A4) reduces to Equation (A3) by employing the following relationships

$$p(\mathsf{R})\, J_2 \prod_{j=1}^{p} p(\mathbb{d}_j|\mathsf{R}) \propto |\Sigma|^{-[\nu_0+p+1]/2} \exp[-\mathrm{tr}(A\Sigma^{-1})/2]$$

$$p(\mathring{\alpha}|\mathsf{R}) \prod_{j=1}^{p} \mathbb{d}_j^{-q/2} \propto |\Sigma|^{-q/2} \exp[-\mathrm{tr}(M\alpha'\Sigma^{-1}\alpha)/2]$$

$$N(\mathring{\mathbf{c}}_j|\mathbb{d}_j^{-1/2}\mathbf{c}_{j0}, \mathbb{d}_j^{-1}V_{cj})\, \mathbb{d}_j^{-(K-2)/2} \propto N(\mathbf{c}_j|\mathbf{c}_{j0}, V_{cj})$$

$$N(\mathring{\mathbf{y}}_{i\mathrm{m}}|\mathbf{x}_i\mathring{\alpha}_i, \mathsf{R}_i) \prod_{j=1}^{s_i} \mathbb{d}_j^{-1/2} \propto N(\mathbf{y}_{i\mathrm{m}}|\mathbf{x}_i\alpha_i, \Sigma_i)$$

$$I(0 < c_{j2} < \dots < c_{j,K-1}) \prod_{i=1}^{n_j} I(y_{ij} \in C_{ij}) = \prod_{k=1}^{K-1} I(\underline{\mathsf{m}}_{jk} \le c_{jk} < \bar{\mathsf{m}}_{jk})$$

where $\mathrm{tr}(A\Sigma^{-1}) = \sum_{j=1}^{p} a_j \mathsf{R}^{jj}/\mathbb{d}_j$ and $c_{j1} = 0$.

While our algorithm specifies the prior on $\mathbf{c}_j$, we can alternatively place a prior directly on $\mathring{\mathbf{c}}_j$. If a flat prior is assigned to $\mathring{\mathbf{c}}_j$, we get the posterior $\pi(\alpha, \Sigma|Y_{\mathrm{m}}, W_o) \propto \psi(\Sigma, \alpha|\nu_0, A, \mathbf{0}, M) \prod_{j=1}^{p} \mathbb{d}_j^{(K-2)/2}$ for $\psi(\cdot)$ defined in Equation (14). The posterior $\mathbf{c}_j|\alpha, \Sigma, Y_{\mathrm{m}}, W_o$ is still given by Equation (16).

Alternatively, if a diffuse normal prior $N(\mathring{\mathbf{c}}_0, \mathring{V}_{cj})$ is put on $\mathring{\mathbf{c}}_j$, the posterior is $\pi(\alpha, \Sigma, \mathbf{c}_j|Y_{\mathrm{m}}, W_o) \propto \psi(\Sigma, \alpha|\nu_0, A, \mathbf{0}, M) \prod_{j=1}^{p} [\mathbb{d}_j^{(K-2)/2} \exp[-(\mathbf{c}_j - \mathbb{d}_j^{1/2}\mathring{\mathbf{c}}_0)'(\mathbb{d}_j\mathring{V}_{cj})^{-1}(\mathbf{c}_j - \mathbb{d}_j^{1/2}\mathring{\mathbf{c}}_0)/2]$. One approach is to draw from $\pi(\alpha, \Sigma|\mathbf{c}_j, Y_{\mathrm{m}}, W_o)$ and $\pi(\mathbf{c}_j|\alpha, \Sigma, Y_{\mathrm{m}}, W_o)$. It is more efficient to sample $(\Sigma, \alpha, \mathbf{c}_j)$ jointly as a block by sequentially drawing from $\pi(\alpha, \Sigma|Y_{\mathrm{m}}, W_o)$ and $\pi(\mathbf{c}_j|\alpha, \Sigma, Y_{\mathrm{m}}, W_o)$. We have $\pi(\alpha, \Sigma|Y_{\mathrm{m}}, W_o) \propto \psi(\Sigma, \alpha|Y_{\mathrm{m}}, W_o) \prod_{j=1}^{p} [\mathbb{d}_j^{(K-2)/2} B_i]$, where $B_i = \int \exp[-(\mathbf{c}_j - \mathbb{d}_j^{1/2}\mathring{\mathbf{c}}_0)'(\mathbb{d}_j\mathring{V}_{cj})^{-1}(\mathbf{c}_j - \mathbb{d}_j^{1/2}\mathring{\mathbf{c}}_0)/2] d(\mathbf{c}_j)$ is bounded above by 1 and approximately 1. The posterior for $\mathbf{c}_j$ is truncated normal defined in Equation (15) with $(\mathbf{c}_0 = \sqrt{\mathbb{d}_j}\mathring{\mathbf{c}}_0, V_{cj} = \mathbb{d}_j\mathring{V}_{cj})$, which is approximately uniform as in Equation

(16).